\documentclass[11pt,preprint]{aastex}
\usepackage{natbib}
\usepackage{icarusbib}
\usepackage{lastpage}

\newcommand{\simlt}{\lower.5ex\hbox{$\; \buildrel < \over \sim \;$}}

\shorttitle{Explosion of Comet Holmes}
\shortauthors{Reach {\it et al.}}

\begin{document}

\title{Explosion of Comet 17P/Holmes as
revealed by the {\it Spitzer} Space Telescope}

\author{William T. Reach}
\affil{Infrared Processing and Analysis Center, 
MS 220-6, California Institute of Technology,
Pasadena, CA 91125}
\email{reach@ipac.caltech.edu}

\author{Jeremie Vaubaillon}
\affil{Institut de M\'ecanique C\'eleste et de Calcul des \'Eph\'em\'erides, 77 avenue Denfert-Rochereau, 75014 Paris, France}

\author{Carey M. Lisse}
\affil{Johns Hopkins Applied Physics Laboratory, SD/SRE-MP3/E167, 11100 Johns Hopkins Road, Laurel, MD 20723}

\author{Mikel Holloway}
\affil{Holloway Comet Observatory, Van Buren, AR}

\author{Jeonghee Rho}
\affil{Spitzer Science Center, MS 220-6, California Institute of Technology,
Pasadena, CA 91125}

\slugcomment{Manuscript Pages: \pageref{LastPage},
Tables: \ref{masstab}, 
Figures: \ref{comp1892}
}

\noindent {\bf Proposed running head:} Explosion of comet Holmes

\bigskip

\def\extra{
\noindent {\bf Editorial correspondence to:} 

Dr. William T. Reach

Infrared Processing and Analysis Center

MS 220-6

Caltech

Pasadena, CA 91125

Phone: 626-395-8565

Fax: 626-432-7484

E-mail: reach\@ipac.caltech.edu
}
\clearpage

\begin{center}
{\Large\bf Abstract}
\end{center}

An explosion on comet 17P/Holmes occurred on 2007 Oct 23, projecting particulate debris of a wide range of sizes into
the interplanetary medium. We observed the comet using the mid-infrared spectrograph (5--40 $\mu$m),
on 2007 Nov 10 and 2008 Feb 27, and the
imaging photometer (24 and 70 $\mu$m), on 2008 Mar 13,
on board the {\it Spitzer} Space Telescope. The 2007 Nov 10 spectral mapping revealed spatially diffuse emission with detailed mineralogical
features, primarily from small crystalline olivine grains. 
The 2008 Feb 27 spectra,  and the central core of the 2007 Nov 10 spectral map, reveal nearly featureless spectra, due to
much larger grains that were ejected from the nucleus more slowly. 
Optical images were obtained on multiple dates spanning 2007 Oct 27 to 2008 Mar 10 at the Holloway Comet Observatory and
1.5-m telescope at Palomar Observatory.
The images and spectra can be segmented into three components:
(1) a hemispherical shell fully $28'$ on the sky in 2008 Mar, due to the fastest (262 m~s$^{-1}$), smallest (2 $\mu$m) debris, with a mass $1.7\times 10^{12}$ g;
(2) a `blob' or `pseudonucleus' offset from the true nucleus and subtending some $10'$ on the sky, due to intermediate speed (93 m~s$^{-1}$) and size (8 $\mu$m) particles, with a total
mass $2.7\times 10^{12}$ g; and
(3) a `core' centered on the nucleus due to slower (9 m~s$^{-1}$), larger (200 $\mu$m) ejecta, with a total mass 
$3.9\times 10^{12}$ g. 
This decomposition of the mid-infrared observations can also explain the temporal evolution of the mm-wave flux.
The orientation of the leading edge of the ejecta shell and the ejecta `blob,'
relative to the nucleus, do not change as the orientation of the Sun changes; instead, the configuration was imprinted by
the orientation of the initial explosion.
The distribution and speed of ejecta implies an explosion in a conical pattern directed approximately
in the solar direction on the date of explosion. 
The kinetic energy of the ejecta $>10^{21}$ erg is greater than the gravitational binding energy of the nucleus.
We model the explosion as being due to crystallization and  release of volatiles from  interior amorphous ice  within
a subsurface cavity; once the pressure in the cavity exceeded the surface strength, the material above the cavity was
propelled from the comet. The size of the cavity and the tensile strength of the upper layer of the nucleus 
are constrained by the observed properties of the ejecta; tensile strengths on $> 10$ m scale must be 
greater than 10 kPa (or else the ejecta energy exceeds the binding energy of the nucleus) and they are plausibly 200 kPa.
The appearance of the 2007 outburst is similar to that witnessed in 1892, but the 1892 explosion was less energetic by
a factor of about 20.



\clearpage

\section{Introduction}

In a dramatic encore to its 1892 explosion, when Edwin Holmes noted a comet that rapidly increased
from obscurity to the brightness of M 31 \citep{barnard13}, comet 17P/Holmes again abruptly increased in
brightness to rivaling the brightest stars in the sky on 2007 Oct 23.
The earliest images of the 2007 event were obtained serendipitously by the SuperWASP-North 
facility, where an image on 2007 Oct 23.27 found the comet at $V\sim 15$ while it saturated the
camera on 2007 Oct 24.10 \citep{hsiehHolmes07}.

The 2007 event was witnessed by an astounded world population (both from dark sites and
suburbs), a large number of amateur
astronomers (including many with CCD cameras), and several professional observers. Some published
professional results include optical images and modeling by \citet{moreno08} and gas chemistry by
\citet{bockeleemorvan08}. The H$_2$O production rate, $Q_{{\rm H}_2{\rm O}}$, was measurements by 
\citet{dellorusso08}, who showed it to decrease from $4.5\times 10^{29}$ to $0.66\times 10^{29}$ s$^{-1}$
from 2007 Oct 27.6 to 2007 Nov 2.3, and \citet{schleicher09}, who measured $5\times 10^{29}$ s$^{-1}$ one week
after the explosion and a decrease by a factor of more than 200 after 125 days. Since the H$_2$O
measurements are based on aperture photometry, 
and the lifetime of OH (the daughter species observed as proxy for H$_2$O) is 14 days at $r=2.5$ AU 
\citep{cochran93},
elevated values inferred for $Q_{{\rm H}_2{\rm O}}$ after the explosion are likely due to enhanced (above normal) but
rapidly decreasing sublimation from the nucleus \citep{schleicher09}.

The 2007 event provided us a new look at a rare and extreme event that may affect many comets during their
lifetimes. The driving mechanism for the 1892 and 2007 eruptions from comet Holmes is not understood, and
modern techniques allow improved spectroscopic and imaging techniques to be brought to bear on the ejecta
from the 2007 explosion. The existence of an energy source within Holmes suggests other comets, and possibly
volatile-bearing asteroids, may also harbor regions of explosive potential. 
The ability of the nucleus to survive explosions with such great energy also indicates it has significant
tensile strength. Regardless of the cause of the explosion, the liberation of a vast cloud of ejecta allows a rare glimpse
of material from the interior of the cometary nucleus. The interior and surface of cometary nuclei may have
different properties due to surface irradiation, depletion of volatiles during perihelion passages, and 
buildup of a mantle of particles larger than $\sim 1$ cm that cannot be levitated from the surface by
gas drag \citep{jewitt05}. 

In this paper we describe infrared imaging and spectroscopic observations with the 
{\it Spitzer} Space Telescope \citep{werner},  and optical imaging with 
1.5-m telescope at Palomar Observatory and the Holloway Comet Observatory.
A dynamical model is developed to explain the images, and a mineralogical model is developed to explain
the spectra. The results are combined and used to determine a possible origin for the 2007 and 1892 explosions based
on heating of subsurface pockets of amorphous ice, which cause subsurface cavities to build pressure until
they rupture the comet's surface.

\section{Observations}

\subsection{Mid-infrared spectra}
Table~\ref{obslog} summarizes the mid-infrared observations made with {\it Spitzer}.
The spectra were taken with the Infrared Spectrograph \citep[IRS;][]{houck04}.
In 2007 Nov and 2008 Feb, the same suite of IRS observations was constructed.
For each subslit of the IRS, a spectral map was performed, and 
a nearby background field was observed. 
Figure~\ref{mips24zoom} (see inset)
shows the sizes of the various spectral maps. 
The spectral maps were combined into cubes using CUBISM \citep{smith07}, including subtracting the background
reference observations, removing bad pixels, and robustly combining the unflagged pixels so as to further
remove time-variable or `rogue' pixels. 
Peak-up images with the 16 and 22 $\mu$m cameras
were taken on both IRS observing dates, but the 2007 Nov peakup images were completely saturated;
Figure~\ref{peakup} shows the peakup images from 2008 Feb.

\begin{figure}
\plotone{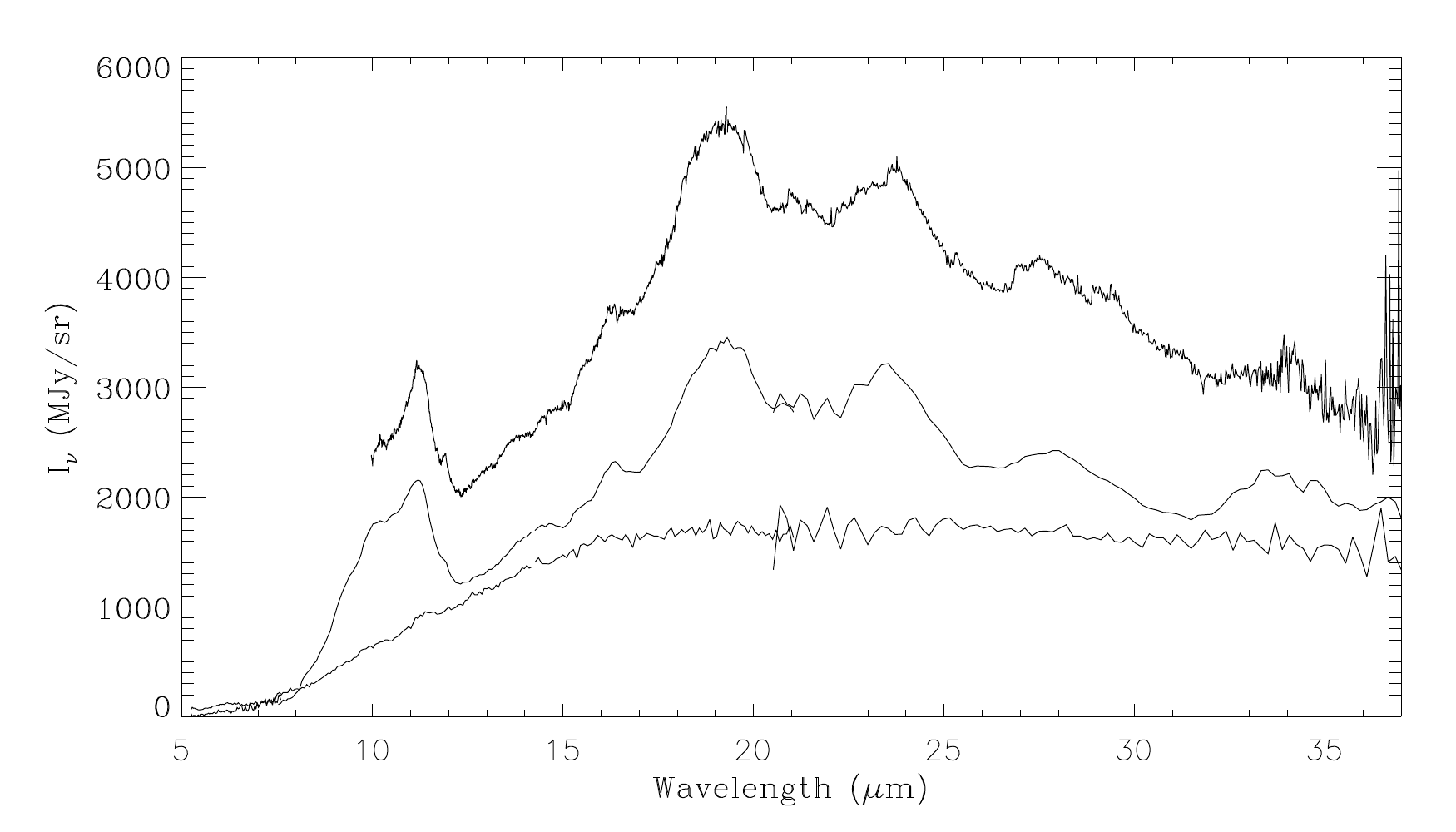}
\figcaption{
Extractions through the {\it Spitzer}/IRS spectral cubes in 2007 Nov.
The upper curve shows the high-spectral-resolution 9.9--37.2 $\mu$m
spectrum of 
the central $6\arcsec$ square region centered on the nucleus.
The middle curve shows the low-resolution 5.2--38 $\mu$m spectrum of the 
`off-center' extraction taken $20\arcsec\times 13\arcsec$  offset $48''$ from the nucleus.
The lower curve is the difference between the low-resolution spectrum of the
``center'' and the low-resolution spectrum of the ``off-center'' region.
If the data can be approximated as a combination of a diffuse, spatially uniform
component (such as the dust shell, which subtends an area much larger than
was covered by the spectrograph) and a central peak, then the lower curve
isolates the emission from the central region in three dimensions.
\label{stitchplot}}
\end{figure}

The brilliance of the comet allowed very high-quality spectral cubes to
be constructed, with significant emission throughout the observed region. 
Figure~\ref{stitchplot} shows spectral extractions.
Using the low-resolution ($\lambda/\Delta\lambda=60$ to 120) data, spectra at all
wavelengths were extracted from two regions. 
The ``central'' extractions cover the $6\arcsec\times 6\arcsec$ ($15\arcsec\times15\arcsec$ for wavelengths longer than 14 $\mu$m) square region centered on the nucleus; the size was chosen to enclose the point-spread function of
the telescope.
The ``off-center'' extractions cover a $20\arcsec\times 11\arcsec$ rectangular region, offset by
$48\arcsec$ from the nucleus (toward position angle 250$^\circ$ E of N) such that it does not overlap with the `central' extraction.

The spectra from the 2007 Nov 10 observations (Fig.~\ref{stitchplot})
reveal distinct, structured silicate emission features. The spectra vary significantly
across the observed region. The main spatial variation is manifested as a distinction
between spectra toward the nucleus and spectra away from the nucleus. The lower curve
in Figure~\ref{stitchplot} shows the difference between the nuclear and off-nuclear spectra,
revealing a near-featureless spectrum. Thus we can approximate, to first order, the emission
as the sum of two components: a spatially uniform component with high-amplitude silicate
emission features, and a spatially compact component that is spectrally featureless.
The spectrally-structured component cannot be spatially concentrated toward the nucleus; 
if it were, then the spatially compact emission would require {\it negative} spectral features at
the locations of the silicate peaks. Such absorptions are not expected considering the small
amount of material present (with line of sight optical depths much less than 1).
The compact component not only lacks silicate features, but it also is relatively brighter at longer
wavelength; which is to say, it has a lower color temperature. Both the lack of spectral features
and the lower color temperature suggest the emission closer to the nucleus is from particles larger
than the wavelength (radius larger than 6 $\mu$m). The high amplitude of the
spectral features away from the nucleus indicate the particles there are smaller than the
wavelength. (The particle composition and size are elaborated in following sections of the
paper.) 

The compact emission centered on the nucleus in 2007 Nov does not arise from the nucleus itself; rather
it must arise from dust. To estimate the nuclear contribution, we use a standard thermal model
and the known diameter \citep[3.4 km][]{LamyNuc} to find the predicted nuclear flux is 0.02 Jy at
24 $\mu$m wavelength. Averaging over the extraction aperture, the predicted 
surface brightness of the nucleus is 350 times fainter than the lower curve in
Figure~\ref{stitchplot}. The predicted spectrum of the bare nucleus is similar to the
observed compact emission. Thus the compact emission must arise from large particles
with a total surface area approximately 350 times that of the nucleus.
At the time of the 2008 Feb observations, the nucleus is predicted to have a flux of 0.012 Jy at
24 $\mu$m; for the extraction in a 15$''$ square centered on the nucleus, we estimate the nucleus is 43 times
fainter than the dust emission.

\subsection{Mid-infrared images\label{mirimagesec}}
\begin{figure}
\includegraphics[width=5in]{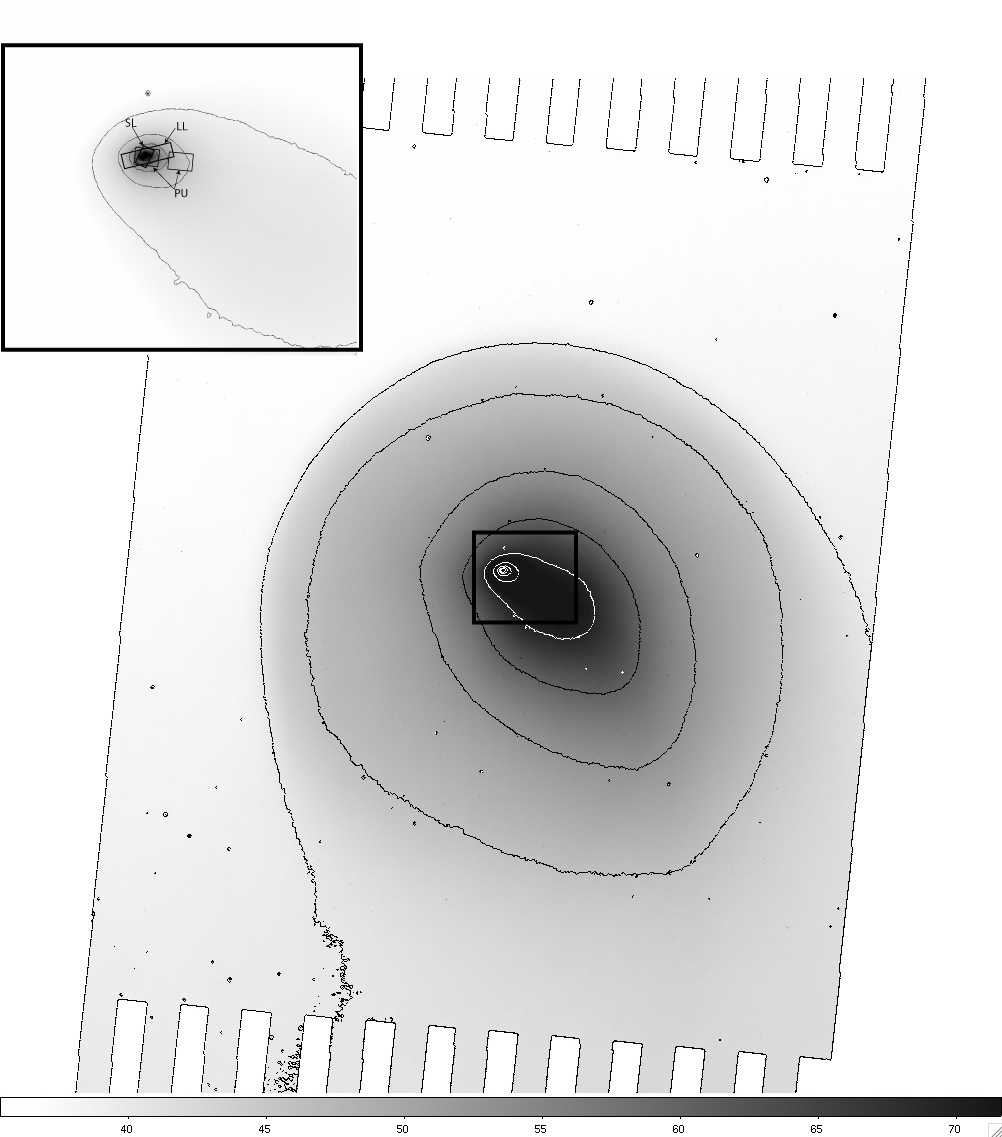}
\figcaption{
MIPS 24 $\mu$m image taken on 2008 Mar 13. 
The black contours are at 35, 40, 46, 52.8, 60.5, and the 
white contours are at 69.4, 79.6, 91.3, 105, and 120 MJy~sr$^{-1}$; the greyscale bar at the bottom is
in MJy~sr$^{-1}$ units. North is up and east is to the left. 
The inset shows a zoom into the region outlined by the black box, $20'\times 17'$ in size, with the image rescaled so as
not to saturate and the contours reversed in color. Small rectangles in the inset show the locations of the
IRS observations from 2008 Feb 27: SL=short-low spectral map, LL=long-low spectral map, PU=blue peakup image.
The same set of IRS observations was performed on 2007 Nov 10.
\label{mips24zoom}}
\end{figure}

\begin{figure}
\epsscale{.5}
\plotone{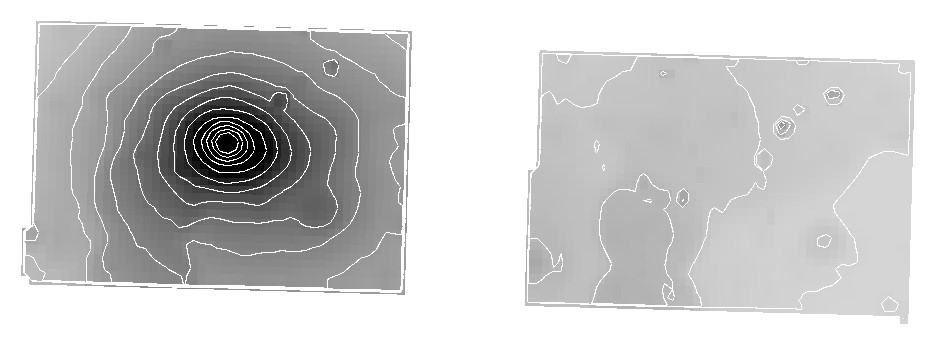}
\plotone{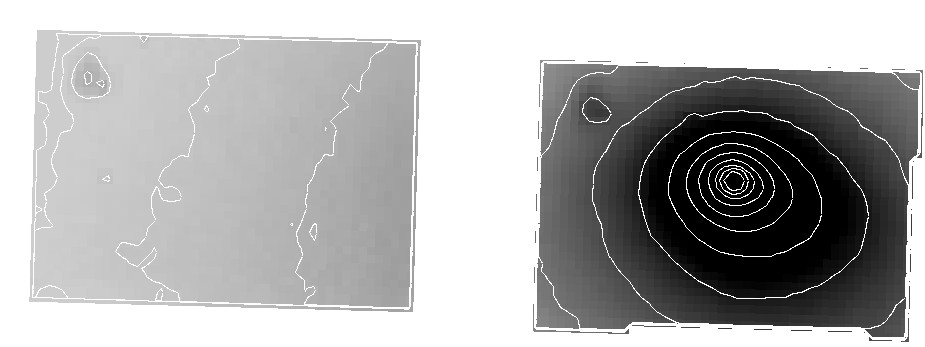}
\epsscale{1}
\figcaption{IRS peak-up images taken on 2008 Feb 27 through the blue (16 $\mu$m wavelength) and red
(22 $\mu$m) filters. The top two rectangles are the blue images, with the left image centered on the nucleus and the right 
(serendipitous) image 
located at its appropriate location on the sky relative to the nucleus.
Similarly the bottom two rectangles are the red images, but this time the nucleus is in the right-hand image and the serendipitous
field is to the left.
The scales are logarithmic, with contours ranging from 34--160 MJy~sr$^{-1}$
in the blue band and 70--270  MJy~sr$^{-1}$ in the red band. 
The size of each rectangular area is $1.41'\times1.03'$.
A few compact sources (likely asteroids and galaxies) are evident in both images.
Extended emission from the comet is detected throughout these images.
The location of the blue peakup images is shown on the wide-field view of the comet in 
Figure~\ref{mips24zoom}. 
Each panel is oriented with North up and East left.
\label{peakup}}
\end{figure}

\begin{figure}
\epsscale{1}
\plottwo{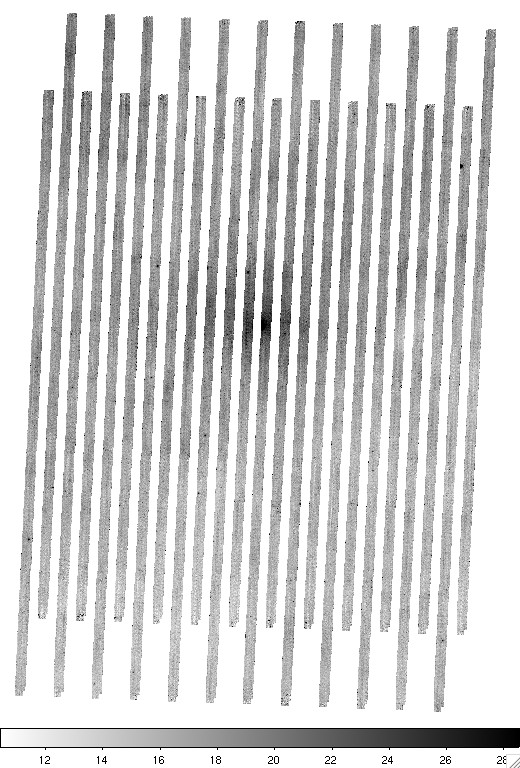}{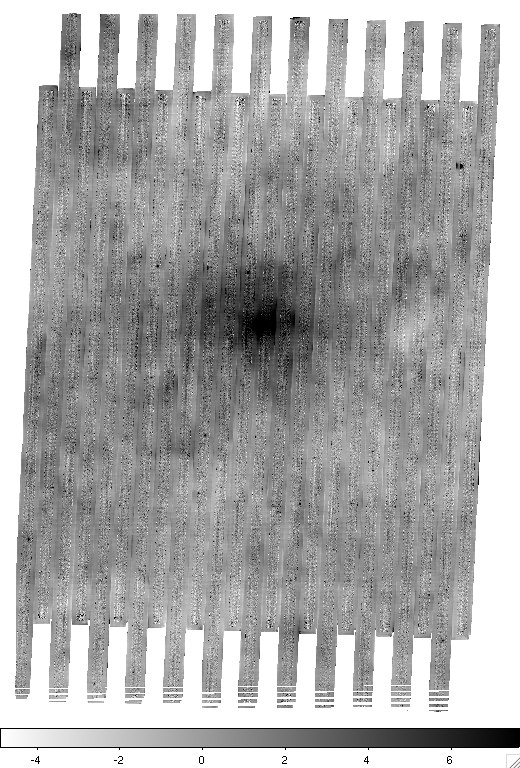}
\epsscale{1}
\figcaption{MIPS 70 $\mu$m images taken on 2008 Mar 13.
The image at left is the 70 $\mu$m mosaic showing all observed data. The sky is half-filled in
a ``picket fence'' pattern, because the width of the functioning portion of the 70 $\mu$m array is half the 
spacing between scans.
The image at right has had the empty spaces filled using a running median, and  background subtracted using the median
in-scan profile (avoiding the $10'$ region around the nucleus). Quantitative work was done only using panel (a), but panel (b) 
better demonstrates the transition (around $9'$ from the nucleus) 
from comet-dominated emission to the patchy background of interstellar dust emission. 
Each panel is oriented with North up and East left. 
Each scan leg is 161$^\prime$ long and the gaps between scan legs are 
$2.5^\prime$ wide. In the image at right, the rectangular area of contiguous data is 
$121^\prime\times 143^\prime$ in size.
The scalebar at the bottom indicates the surface brightness in MJy~sr$^{-1}$; it differs between
panels primarily due to background subtraction when constructing the filled mosaic.
\label{mips70}}
\end{figure}

On 2008 Mar 13, a wide-field image, spanning $2^\circ\times 2^\circ$, 
was created with the Multiband Infrared Photometer for {\it Spitzer} \citep[MIPS;][]{rieke04}.
Figure~\ref{mips24zoom} shows the 24 $\mu$m MIPS image, together with the locations of the various IRS observations.
At the same time as the 24 $\mu$m array was observing, a 70 $\mu$m image was created. The operating portion
of the 70 $\mu$m array is only half as wide as that of the 24 $\mu$m array, in the in-scan direction, so the 
70 $\mu$m image is only half filled, as evident in Figure~\ref{mips70}a.
The MIPS observing sequence was repeated on 2008 Oct 24, but using the same celestial coordinates as the comet
had in 2008 Mar. This `shadow' observation was used to trace the celestial background, due to thermal emission from interstellar dust. The empty portions of the primary and shadow images were 
filled using a median over a $7.7'$ square, as in Figure~\ref{mips70}b, then subtracted. 

\begin{figure}
\epsscale{.7}
\plotone{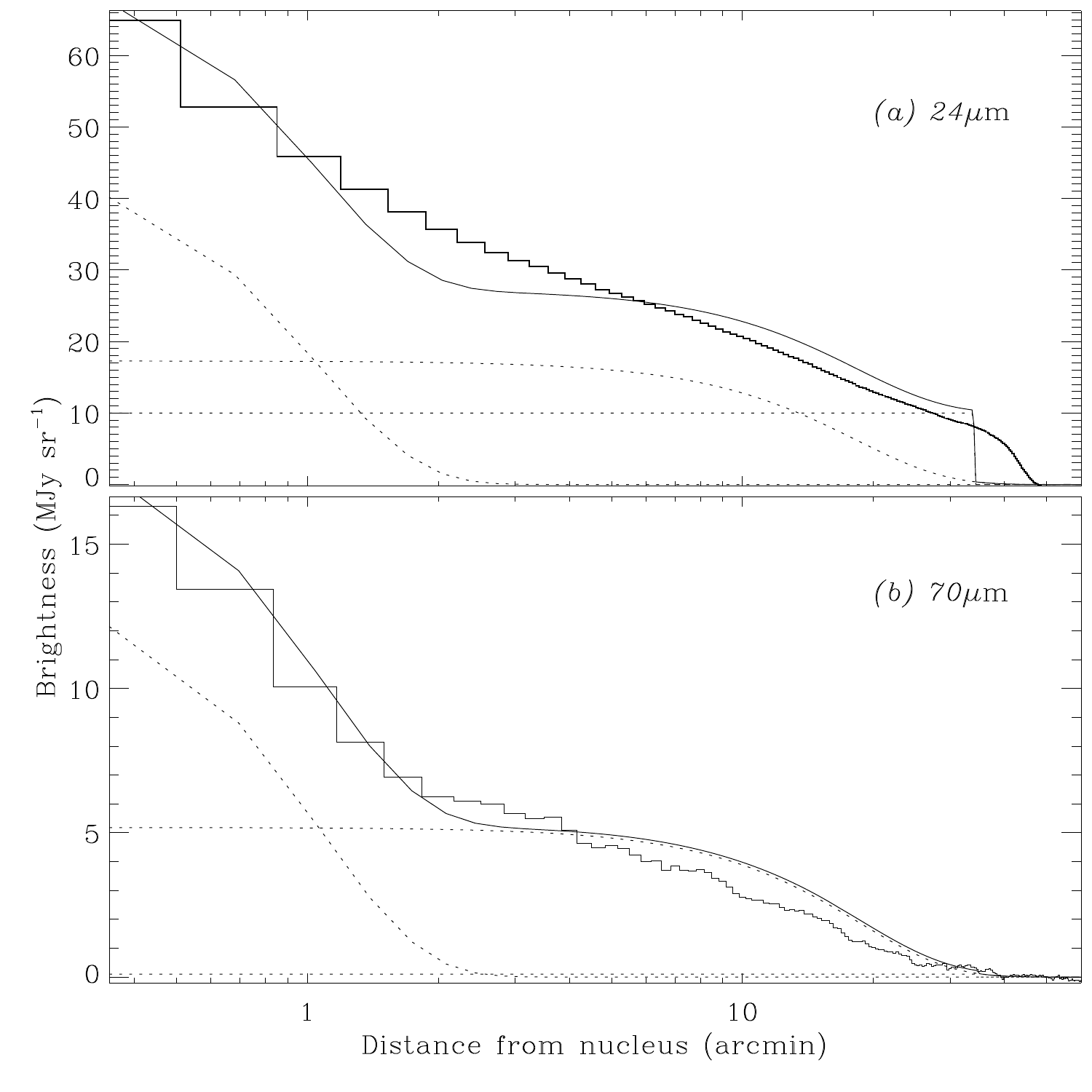}
\epsscale{1}
\figcaption{
Radial (azimuthally-averaged) surface brightness profiles through the shadow-subtracted {\it Spitzer}/MIPS images at 
{\it (a)} 24 $\mu$m and {\it (b)} 70 $\mu$m.
The radial profiles through the images are stair-stepped lines, while three components
(core, blob, shell) are a set of dotted curves, and the sum of the components is a thin, solid curve. 
The shell is not detected at 70 $\mu$m.
\label{radprof}}
\end{figure}

We modeled the MIPS images using a sum of three spatial components: a gaussian core centered on the nucleus with FWHM 1$^\prime$, a gaussian `blob' centered 9$^\prime$ anti-sunward of the nucleus with FWHM 13$^\prime$, and a circular region of constant brightness with radius 35$^\prime$ that crudely approximates a `shell.'
The significance of this decomposition is discussed in \S\ref{massenergy}.
Figure~\ref{radprof} shows radial profiles through the 24 and 70 $\mu$m shadow-subtracted images and the model components. 
The core and blob components of the comet are present at both wavelengths, with a brightness ratio
70/24 $\mu$m of 0.3. 
The outermost component in the 24 $\mu$m image, which corresponds to the fast-moving shell seen in the early optical images, is much fainter at 70 $\mu$m, with an estimated
brightness ratio of 70/24 $\mu$m of less than 0.03.
For a greybody spectrum, the 70/24 $\mu$m brightness ratio indicates a color temperature 
of 260 K for the core and blob, while the shell cannot be explained.
But for small grains with emissivity proportional to frequency squared, the shell can be
explained by a temperature greater than or equal to 400 K. We will return to particle size
and temperature estimates when interpreting the spectrum, but it is worth noting that with
the complete spatial coverage and high resolution of the MIPS images, it is already possible
to conclude that the shell is due to small particles (radius less than $\lambda/2\pi$,
which for 24 $\mu$m image means radius less than 4 $\mu$m)
while the core and blob can be explained by larger particles 
(radius greater than 11 $\mu$m to emit efficiently at 70 $\mu$m).

\subsection{Optical images}
Optical images were obtained on multiple dates spanning 2007 Oct 27 to 2008 Mar 10 at the Holloway Comet Observatory
near Van Buren, Arkansas (94.35$^\circ$ W, 94.35$^\circ$ N).
Table~\ref{obslogopt} lists the dates and observing circumstances.
The telescope is a TeleVue NP-127is, a 12.7 cm refractor with a fast f5.2 focal ratio; it is mounted on 
a Losmandy G-11 equatorial mount. The images were made using an SBIG ST-10XMEI CCD camera with 
2184$\times$1472 pixels. The pixel scale was increased from $2.1''$ for the early images (total field of view $1.3^\circ\times 0.9^\circ$)
to $11.3''$ in 2008 Mar (total field of view $6.5^\circ\times 4.4^\circ$) as the comet grew.
The precise locations and orientations of the images were determined using the {\tt astrometry.net} tools \citep{hoggastrom,langthesis}.

A deep optical mosaic was obtained on 2007 Nov 06 using the
1.5-m telescope at Palomar Observatory. The images were made using a 2048$\times 2048$ pixel CCD with $0.38''$ 
pixel field of view. 
The telescope is robotically controlled and queue scheduled. Wildfires were raging in the valley below the observatory
on the day of the comet's outburst; our observations were the first ones performed at Palomar after the observatory
reopened. 
A pattern of 7 exposures (2 min each) spanning from far W to far E of the comet was repeated 7 times with small offsets in declination.
In this way each location was observed 7 times, with an interval of 17 min between exposures.
The images were combined in the comet's rest frame
into a mosaic using {\tt mopex} \citep{makovozproj,makovozoverlap}, which matched the background in overlapping
portions of images and combined the images using a dual outlier-rejection algorithm. In this way the stars and galaxies were
almost completely removed, since the motion of the comet was $7.8''$ (much larger than the seeing, which was $1.8''$) 
between images of a given sky patch.

\clearpage

\section{Spectrum in 2007 November: amorphous and crystalline dust}

\begin{figure}
\epsscale{.8}
\plotone{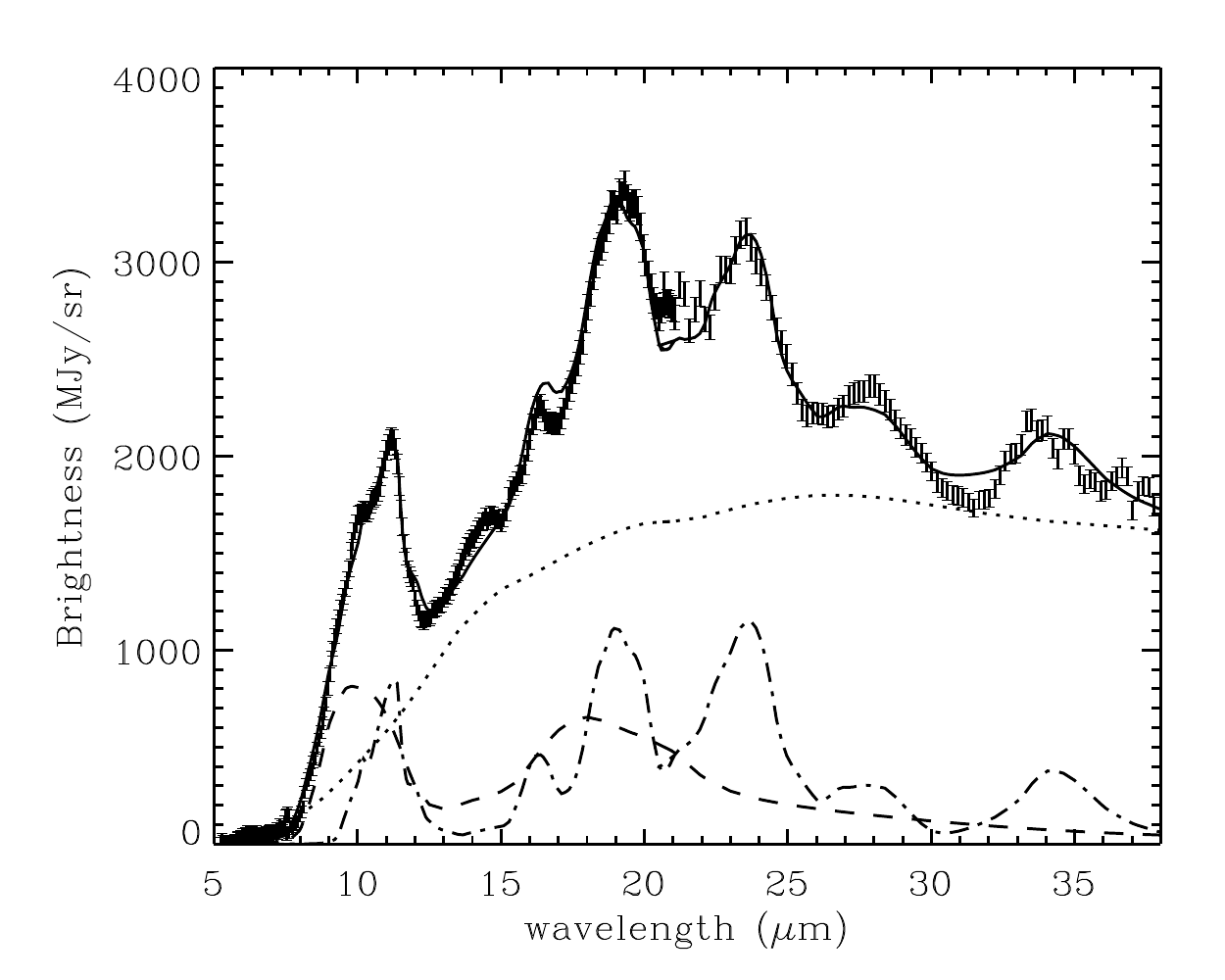}
\epsscale{1}
\figcaption{Spectrum of comet Holmes in 2007 Nov (points with error bars) together
with a 3 component fit: large amorphous pyroxene (dotted line), small amorphous
pyroxene grains (dashed line), and small crystalline forsterite grains
(dash-dotted line).
\label{stitchfit}}
\end{figure}

The spectrum of the diffuse emission, observed soon after the explosion in 2007 November, can
be explained by a combination of amorphous and crystalline
silicate materials. 
Figure~\ref{stitchfit} shows the spectrum of a diffuse region
of the ejecta, near but {\it not} including the nucleus.
As a first attempt to determine the prominent minerals, we made
a simple, three-component fit. One component, chosen based on the locations
of the prominent spectral emission features, is small grains
of crystalline Mg-rich olivine (forsterite; Mg$_{2}$SiO$_{4}$); 
optical constants were taken from \citet{jaeger03}.
Other silicate compositions including enstatite, bronzite, and
montmorillonite, failed to present the prominent features (but
may be present at lower abundance). The spectral shape of the
crystalline silicate model is a blackbody times the absorption efficiency for
a continuous distribution of ellipsoids (CDE) of forsterite;
the temperature is determined by fit.
The second component, chosen for its broad 9--11 $\mu$m 
emission feature, is small grains or amorphous silicate; 
we used the optical constants of amorphous olivine and pyroxene
from \citet{dorschner95}. The spectral shape is a blackbody times
the absorption efficiency
from Mie theory (for spheres calculated from the index of refraction and
for a range of sizes); the temperature and particle size are determined by fit.
The third component must be relatively featureless in spectral shape; we
take it here to be large grains of the same amorphous composition, with their
temperature and size determined by fit.
The temperatures and abundances of the three components,
and the sizes of the small and large amorphous grains,
were fit simultaneously 
using the {\tt mpfit} algorithm of C. Markwardt.
The temperatures for the forsterite, small amorphous, and large amorphous grains
were 223, 309, and 221 K, respectively.
The forsterite grains are in the small-particle limit, and the model
radii of the small and large olivine grains were 1 and 12 $\mu$m.
The quality of this simple fit can be judged from Fig.~\ref{stitchfit};
the formal $\chi^2/\nu$ is 2.76.
The locations of the spectral features are reasonably well
matched, though their amplitudes are not. Given the high quality of
the data, we are compelled to use a more sophisticated model for the
spectrum incorporating more sizes and compositions.

\begin{figure}
\epsscale{.5}
\plotone{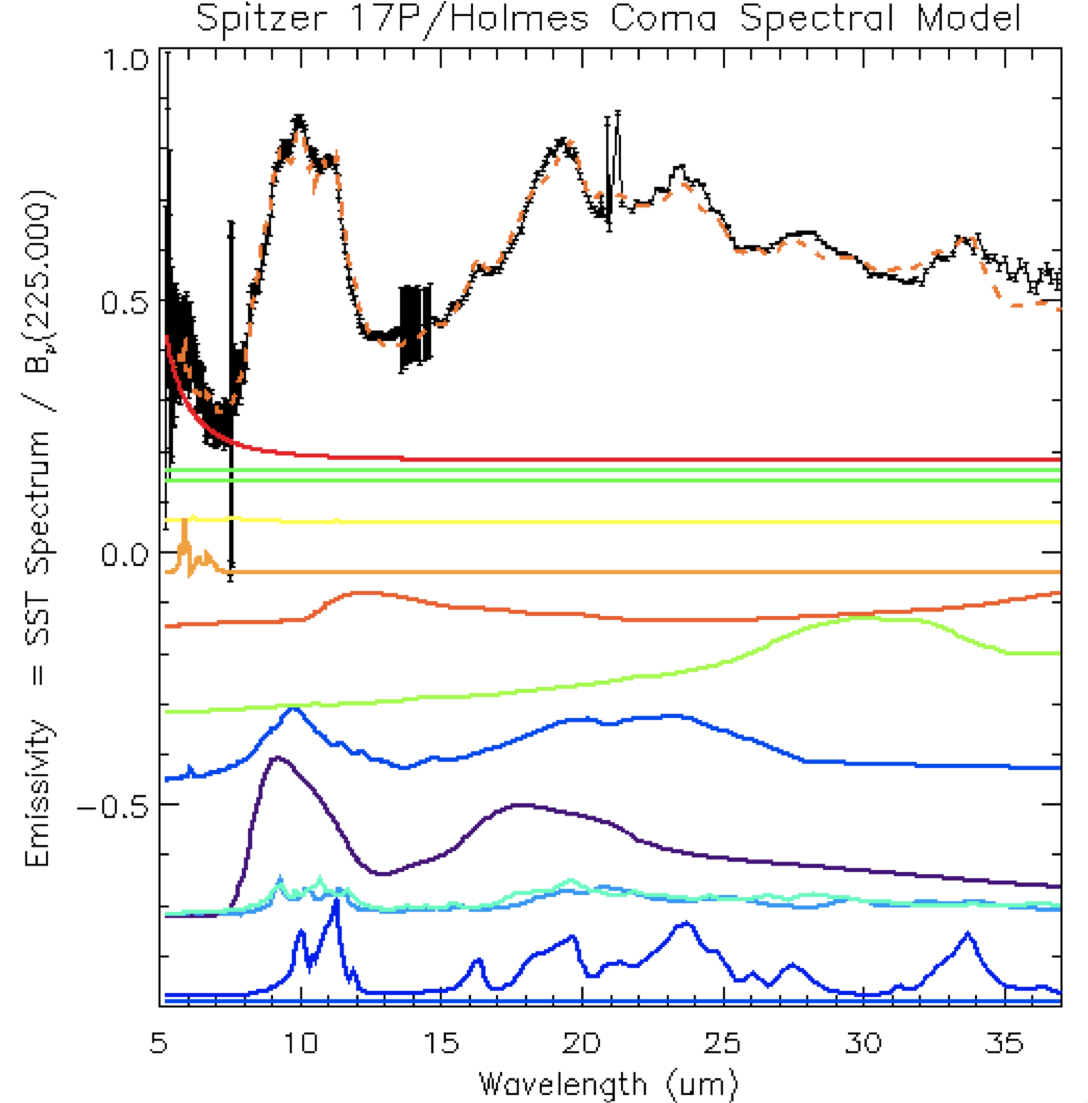} 
\plotone{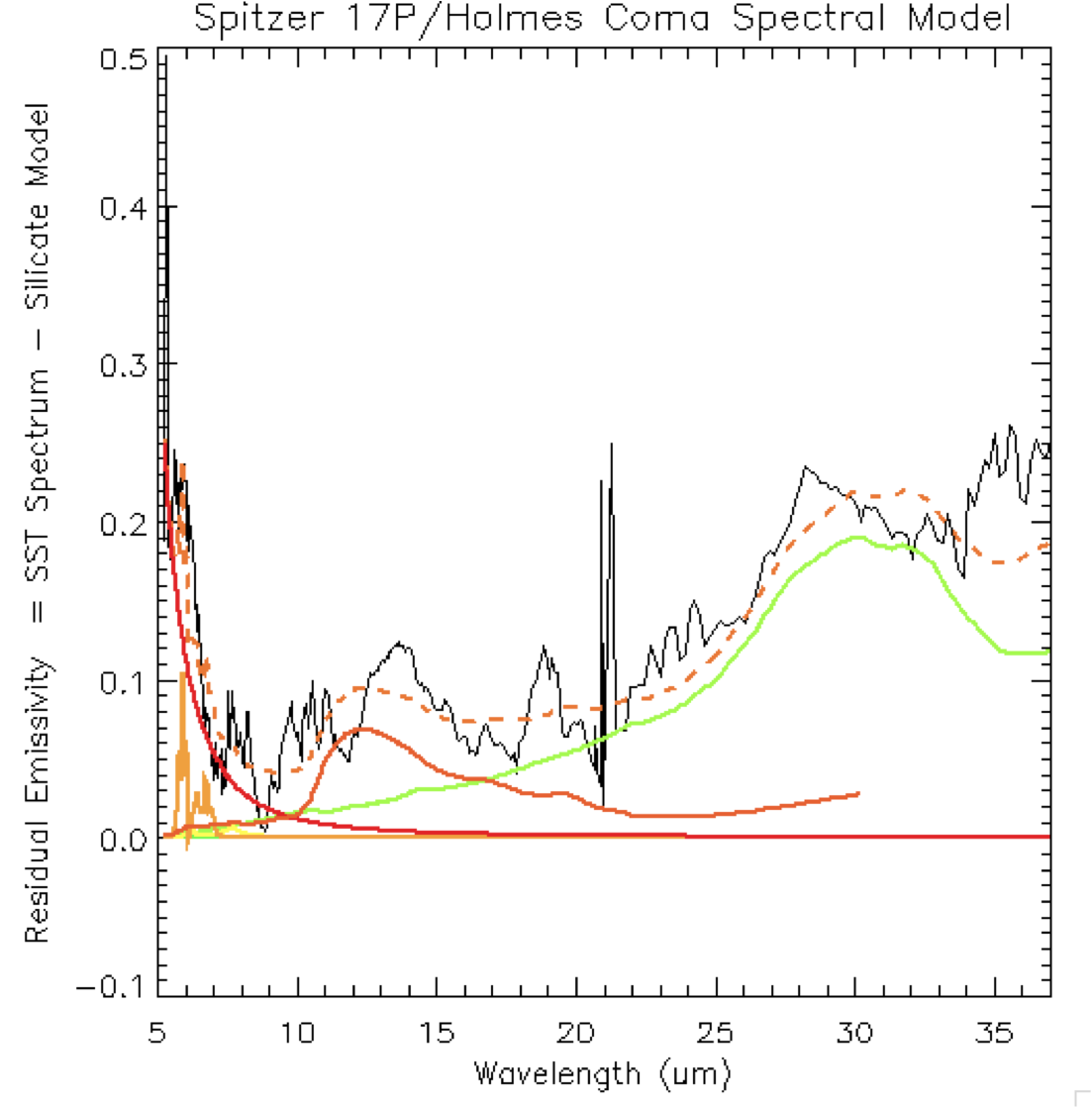}
\epsscale{1}
\figcaption{Emissivity spectrum of comet Holmes in 2007 Nov together
with a mineralogical fit. The top panel shows the fit (orange) together with the
individual mineral constituents (offset vertically for clarity). The bottom panel shows
the fit after subtracting the silicate minerals, so that the sulfide and water 
contributions can be assessed.
\label{specmod}}
\end{figure}

The detailed mineralogy of the dust from comet Holmes was determined by comparing to
a wide range of plausible minerals with high-quality mid-infrared transmission or thermal emission
laboratory,
using the same procedures as documented in detail by \citet{lisse06,lisse07}.
Figure~\ref{specmod} shows the observed spectrum and appropriately scaled
contributions from each mineral.
A size distribution with number density per unit grain size $\propto a^{-4}$ was assumed.
Small grain temperatures were 225 K for all materials except water vapor and ice (170 K) 
and amorphous carbon (410 K); large grains were set to an equilibrium temperature of 180 K.
The mass within the $15''$ extraction area is $5\times 10^{10}$ g.
An excellent fit, with reduced $\chi^2/\nu=0.91$ was obtained.
Table~\ref{comptab} lists the minerals included in the fit.
In terms of atomic abundances, the solids ejected by Holmes are similar to solar
abundance (relative to Si) in Mg, Fe, S, Ca, and Al. 
That refractory elements are present in near-solar abundances is similar to what is
seen for other comet dust. 

The abundance of C is 34\% solar, which is higher than seen in the Deep Impact 9P/Tempel 1
ejecta or from C/Hale-Bopp, but not as high as in 73P/Schwassmann 3.
The abundance of O in solids is 14\% solar, but there is likely to be 
more O in gaseous form than solid. The high temperature of amorphous C grains makes
them important for explaining the near-infrared emission, dominating the predicted brightness from
5--7.5 $\mu$m in the models for the {\it Spitzer} spectrum. Ground-based near-infrared observations
revealed thermal emission at 3.2--4.2 $\mu$m with a color temperature $360\pm 40$ K \citep{yang09},
which we associate with the amorphous C seen in the {\it Spitzer} spectra.

\def\extra{
Some of the mineralogical results are intimately related with the size distribution in the model.
In particular, there is a partial degeneracy between large (presumably silicate) particles and 
amorphous carbon. Within the context of the \citet{lisse06} model, which uses a power-law
size distribution, the featureless continuum is provided by amorphous carbon. This same
conclusion was obtained for the coma of 2P/Encke by \citet{kelley06},
who explained the infrared spectrum of the coma using 0.4 $\mu$m amorphous carbon grains.
Such grains are likely to be present from both 17P and 2P; however for 2P we know from
dynamical modeling that the bulk of the surface area arises from large particles \citep{reachEncke}.
A large-grain explanation (as used for Fig.~\ref{stitchfit}) provides a similar spectral shape for 2P/Encke and for 17P, but
it requires a non-power-law size distribution for the grains. We cannot distinguish between
the two hypotheses (small carbon versus large silicate grains) 
using the present spectral data alone.
}

\clearpage

\section{Spectrum in 2008 February: Large particles}

The spectrum comet Holmes in 2008 Feb is 
completely different from that seen soon after the outburst
in 2007 Nov. Instead of a highly structured spectrum with strong
silicate emission features, the 2008 Feb spectrum is 
dominated by a nearly featureless continuum.
The comet was clearly detected at 70 $\mu$m as well as within
the IRS wavelengths, 5--40 $\mu$m.
The comet was not detected at wavelengths smaller than 7 $\mu$m.
The nuclear spectrum was predicted using a standard thermal
model for a body with diameter 3.4 km. The predicted nucleus
brightness is 2\% (4\%) of the observed flux at 70 $\mu$m (10 $\mu$m)
extracted within a box $15''\times 15''$ in size.
The spatial information in the
spectral maps (and the MIPS 24 $\mu$m image) also confirm that
the extracted flux arises from extended emission rather than the nucleus.
The peak surface brightness in an image at 15.6 $\mu$m 
has a predicted nuclear contribution of 19\% at a resolution of $4.6''$,
which is nontrivial, but for the extended emission we consider a
much larger extraction region, within which the nuclear contribution is diluted to
a negligible 2\%.

To model the spectrum, we used amorphous
grains and optical constants taken from \citet{dorschner95}.
The absorption and scattering efficiencies were calculated for
spherical particles from 0.01 to 1000 $\mu$m using Mie theory.
Two sizes of amorphous olivine, pyroxene, and carbon were considered,
with the size, temperature and amount determined by fit.
For single compositions, the best fit was obtained with amorphous
pyroxene ($\chi^2_\nu=3.13$) compared to
amorphous olivine ($\chi^2_\nu=3.09$) or carbon ($\chi^2_\nu=7.32$).
Including small grains of crystalline forsterite did not significantly improve the fit
($\chi^2_\nu=3.12$). Only a single particle size was required; adding a second
particle size did not improve the fit. The size must be greater than 16 $\mu$m radius,
and the best-fit temperature is 205 K.

\begin{figure}
\epsscale{.8}
\plotone{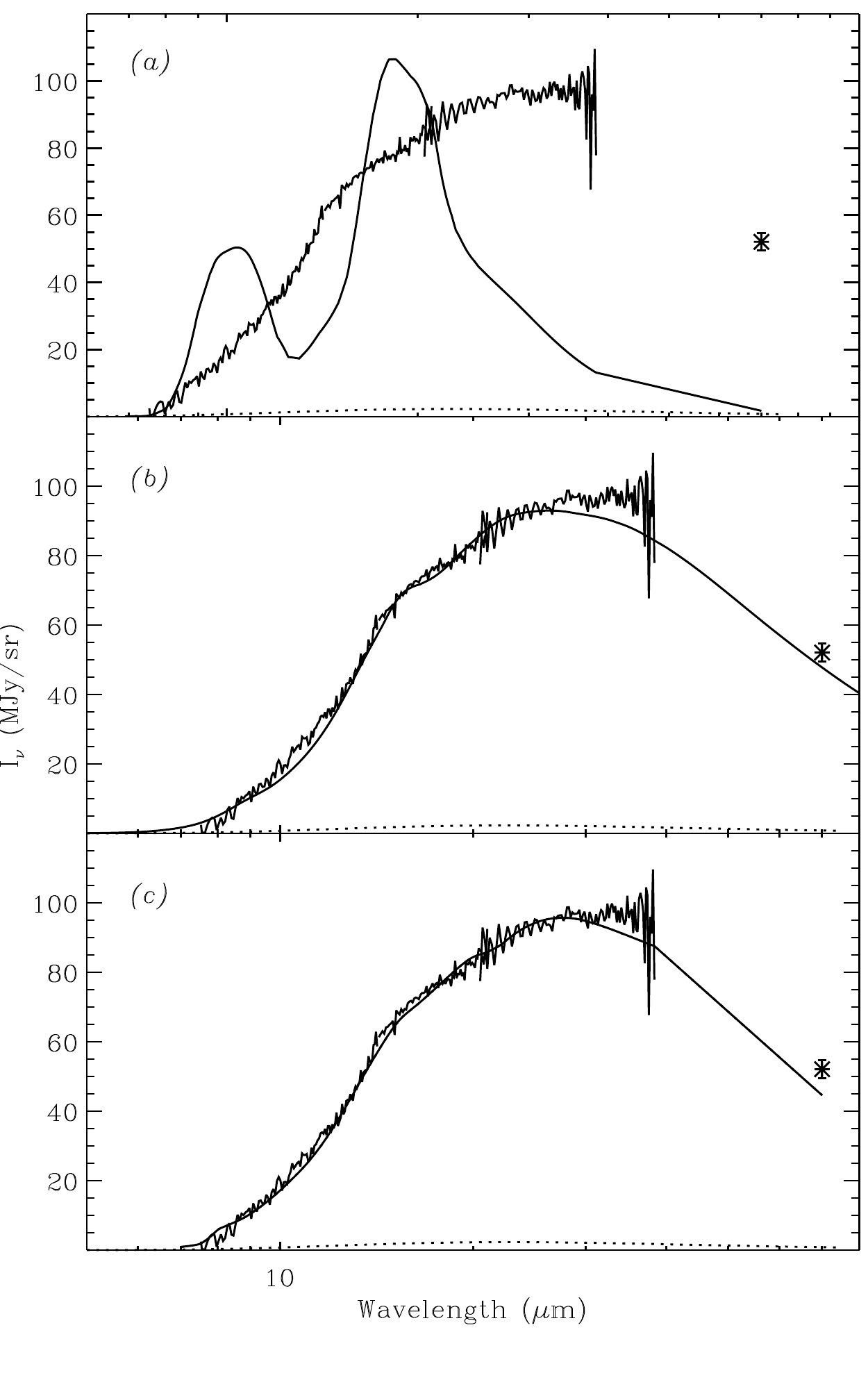}
\epsscale{1}
\figcaption{Spectrum of comet Holmes in 2008 Feb, combined with
the MIPS 70 $\mu$m brightness. The solid lines in each panel
show fits with spherical grains of the following type:
{\it (a)} 3 $\mu$m radius,
{\it (b)} 206 $\mu$m radius,
{\it (c)} 3 $\mu$m and 300 $\mu$m radius.
The estimated flux from the nucleus surface is shown as a dotted line in
each panel.
\label{stitchmar}}
\end{figure}

To illustrate the constraint on small particle contribution to the spectrum,
consider Figure~\ref{stitchmar}a which shows a fit with 
3 $\mu$m grains, at best-fit temperature 150 K. 
This model under-predicts the long-wavelength emission 
by a wide margin and has the prominent 10 and 18 $\mu$m
emission bands that were so important in 2007 Nov but are
absent in 2008 Feb.
Figure~\ref{stitchmar}b shows a fit with 206 $\mu$m grains
at a best-fit temperature of 205 K.
These grains are so large that their spectrum
is not dominated by true emission features
(which are due to the imaginary part of the index of refraction)
but rather by low-amplitude emissivity features 
(combining the real and imaginary parts of the index of refraction).

The composition of the grains is not uniquely determined in 2008 Feb,
although a silicates fit much better than carbon.
It is possible that other grain compositions with nearly featureless
infrared emissivity are present, and it is possible carbon is
present as a minor contributor.
The strongest result is that small silicate grains are absent
from the 2008 Feb spectrum. Their absence is readily explained
by their having been produced in the explosion, from which they rapidly
expanded into a large diffuse cloud (see in the very wide-field MIPS 24 $\mu$m image).
Any small amorphous carbon grains should have done the same thing.
It is natural to expect
larger particles of the same composition seen in the 2007 Nov spectra 
to be present in 2008 Feb, so the fact that amorphous pyroxene can fit
the spectra at both epochs is reassuring.

For large grains, the
emissivity is not strongly dependent on grain size, so
the contribution of grains of a given size 
to the observed brightness is proportional to their surface area,
$dI/da\propto a^{2} dn/da$.
We can make a rough evaluation of the power-law size distribution
index from the fits with two particles sizes, yielding 
$dn/da\propto a^{-\alpha}$ with $\alpha=2.2$.
Such a size distribution is very top heavy, with
most of the surface area in large grains 
and essentially all of the mass in the largest grains,
which is not surprising given that we are observing only
the slowest particles from the explosion.

\section{Properties of the ejecta images}

\subsection{Fine structure: filaments and trail}

\begin{figure}
\epsscale{1}
\plotone{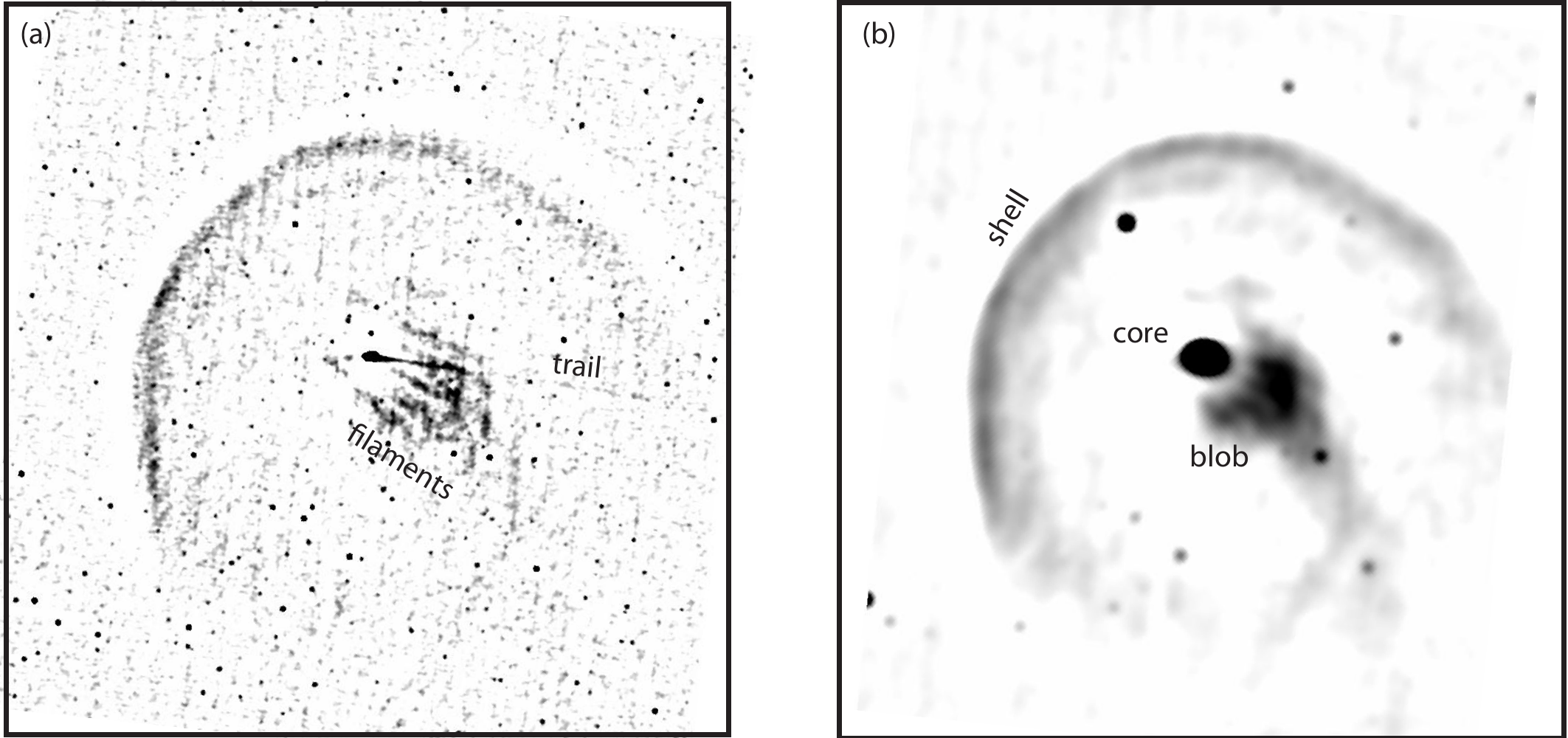}
\epsscale{1}
\figcaption{Wavelet transforms of the 24 $\mu$m image of comet Holmes taken in 2008 Mar from {\it Spitzer}.
 (a) Sum of wavelet scales 3-4, emphasizing coherent structure on scales significantly larger than the pixel-to-pixel
noise.   (b) Sum of wavelet scales 5-6, emphasizing structure on scales larger than shown in panel (a) but still
suppressing the overall smooth distribution of material. The various morphological features that are mentioned in the text
are labeled. In the smoother panel  (b), the hemispherical shell is readily evident, together with the `blob' of emission
located roughly opposite the central `core' which contains the nucleus. In the finer panel (a), the `blob' resolves into 
a set of filaments, and the debris trail that precisely follows the comet's projected orbit becomes evident. 
The images are oriented with celestial N up and E to the left, and the outline boxes are $2^\circ$ on a side.
\label{wave24}}
\end{figure}

\begin{figure}
\epsscale{.6}
\includegraphics{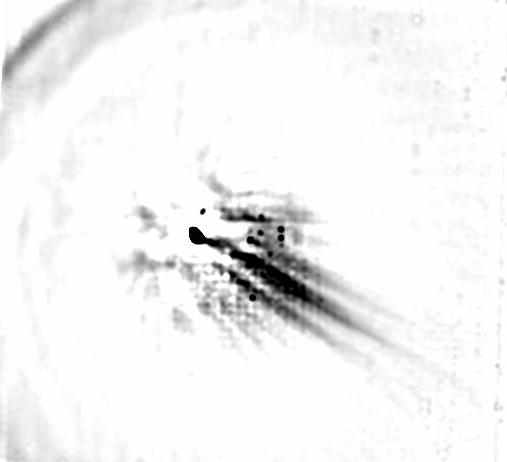}
\epsscale{1}
\figcaption{Wavelet transform of  the R-band image of comet Holmes taken on 2007 Nov 6 at Palomar. 
This is the central portion of the sum of wavelet scales 5-6. The same filamentary structure that is evident
4 months later in the 24 $\mu$m image can be seen in this optical image. Part of the shell due to the
fastest-moving ejecta is evident in the upper-right and lower-left corners of this image. The image is oriented with N up and E to the left,
and it spans $13'$ on a side.
\label{Palwave}}
\end{figure}

The optical and infrared images of the ejecta are fairly smooth, but their signal to noise ratio is so high that we can 
distinguish a wealth of substructure. To highlight the structures, we used a multi-resolution wavelet decomposition,
specifically an {\it \'a trous} decomposition into 8 angular scales \citep{starckAstIm}. This choice of
filters has the great advantage that the original image is the sum of the 8 images, and the image at each wavelet
simply contains the structure at progressively finer angular scale.
The finest scales contain noise and stars, and the coarsest scales contain the overall smooth distribution.
Figure~\ref{wave24} shows the sums over two sets of intermediate angular scales. 
The same processing was applied to the Palomar image; Figure~\ref{Palwave} shows the intermediate scale
transform of the central portion of the image.

The contrast-enhanced images show that the blob of ejecta that was so prominent in the distribution of debris from
comet Holmes contains some 10 separate filaments. The filaments are straight in the images taken soon after the
explosion (Fig.~\ref{Palwave}). But in the images 4 months later (Fig.~\ref{wave24}),
the filaments may connect to curved extensions; these extensions are easier to see in the coarser wavelet (Fig.~\ref{wave24}b).
It is likely that the same exact filaments are present at
the two epochs, but the differing geometries of the lines of sight (from Earth and from Spitzer to the comet, on
two widely different dates) make a detailed cross-identification difficult.

In the 2008 Mar image, a cometary debris trail is evident, running from the nucleus along the projected orbit of the comet.
The properties of this debris trail are typical of those seen in a survey of 35 short-period comets \citep{reachtrail}.
This debris trail is due to mm-sized particles produced on the previous perihelion passes and slowly spreading
out along the comet's orbit.
If the comet had not experienced its 2007 Oct outburst, then the only features that would have been evident in the
image are compact source at the location of the nucleus and the debris trail.

\subsection{Asteroids and fragments}

The MIPS 24 $\mu$m image from 2008 Oct 25 `shadow' covers the same field as the comet image from 2008 Mar 13, except the comet had
long since left the field. Subtracting the shadow image from the comet image, we can assess whether any of the compact sources
are potentially related to the comet (i.e. comoving with the nucleus). 
Most of the sources in the shadow image are stars and galaxies; they are clearly present at both epochs and the fainter ones subtract
accurately. To avoid confusion due to artifacts we masked small regions around the brighter sources in the shadow image.
There are several compact sources in the shadow-subtracted image of the comet. 
To determine which of these are known asteroids, we used the moving object overlay feature in the Spitzer Planning Observation
Tool (SPOT) to overplot locations of all known asteroids on the date of observation. 
Of the 10 known asteroids within $1^\circ$ of the nucleus, the 5 with the brightest predicted visual magnitude were 
clearly detected in the 24 $\mu$m image. These objects (and their flux at 24 $\mu$m) are:
1996 VC11 (151 mJy), 2000 ER111 (215 mJy), Annapavlova (132 mJy), 2000 PF24 (107 mJy), and 2001 QB81 (59 mJy),
and 2009 DZ32 (3.9 mJy).
Other asteroids predicted to be in the field had predicted visible magnitude greater than 20 and were not evident above 
a flux of 5 mJy at 24 $\mu$m. The density of moving objects is typical of what we have seen in other
observations at moderate ($20^\circ$) ecliptic latitude. The Infrared Small Asteroid Model 
\citep[][]{kiss08}
prediction for
our observing location, date, time, and origin ({\it Spitzer}) is 4 asteroids brighter than 5 mJy, and we see 5.

We located all other compact sources in the image above 2 mJy using a spatial high-pass filter.
The three brightest compact sources are located in a nearly straight line to the NE of the nucleus. The fluxes of the three sources are 3.2 mJy each,
and they are likely to be images of the same fast-moving object seen at different positions on different scan legs used to build the map.
The object is not fixed on the array (it moves with the sky as the scan mirror moves between frames) but its
motion during the period when the telescope scanned over the object (4 frames with an timespan of 11 sec 
for each scan leg that crossed the object's position) was not
detected. Thus these three objects could be 3 separate, unknown asteroids, that happen to fall nearly in a line. This would
be an improbable coincidence, so we will explore the possibility they are really the same object.
The trajectory is nearly perpendicular to the projected vector
to the nucleus, so this fast-moving object is not comoving with comet Holmes. 
The arc length is 11.7$'$ between these sightings, which occurred on three scan legs.
From the individual frames we see the object on 2008 Mar 13 at the following UT times (and J2000 RA and Dec):
06:61:06 (04:01:42.3 +40:50:35.0), 07:02:35 (04:02:10.3 +40:47:49.8), 07:12:14 (04:02:35.1 +40:44:25.0).
The angular motion between the first and second sightings is 1900$''$/hr, and between the second and third sightings it
is 2150$''$/hr. The direction of motion between the first and second sightings was 117$^\circ$ E of N, and the
direction of motion between the second and third sightings was 126$^\circ$ E of N, so the arc is slightly curved.
The object should have been covered in the scan legs before and after those with the sightings, but it was not
seen above a level of 2 mJy at the predicted locations, which were calculated by extending the arc before and after 
the observed sightings using
a quadratic fit to the RA and Dec versus time.
The observations were reported to the Minor Planet Center and compared to up-to-date near-Earth object lists, but
no correspondence was found (thanks to G. Williams for checking).
This may be one of the few (or only) near-Earth objects discovered using mid-infrared observations, but 
we can do no more with the available information. 

In terms of possible comet Holmes fragments, we find none above a limit of 5 mJy.
An object with this flux at the location of comet Holmes
would have a diameter of 5 km
according to the standard thermal model with $\eta=0.9$, $\epsilon=0.9$
\citep{lebofsky86}, i.e. as large as comet Holmes itself, so the upper limit on a bare nucleus is not 
very interesting.
But compared to the core of comet Holmes, the upper limit is 3000 times fainter, so the
upper limit does rule out active cometary fragments.
The recent finding by \citet{stevenson09} of small, transient fragments ejected from Holmes demonstrates
that some $\sim 10$ m fragments were produced but were short-lived. Such fragments would be below
our detection limit even if they were still being produced in 2008 Mar.

\subsection{Comparing optical and mid-infrared images}

\begin{figure}
\epsscale{1}
\plotone{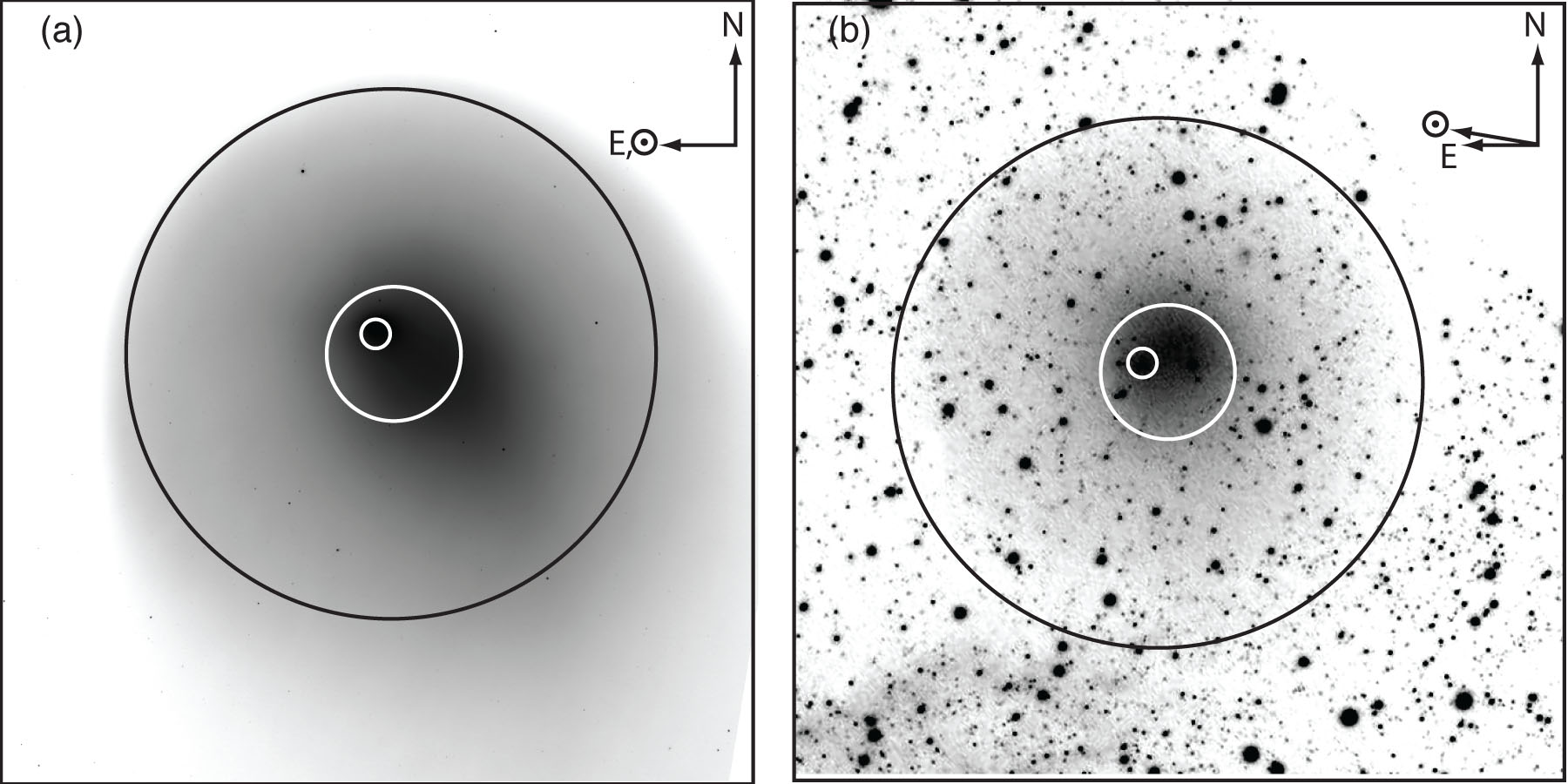}
\epsscale{1}
\figcaption{Comparison of the (a) MIPS 24 $\mu$m image from 2008 Mar 13 to the (b) Holloway Comet Observatory
optical image on 2008 Mar 5. The orientation of each image and the vector to the Sun are labeled. Three circles 
(identical in both panels) are overlaid so as to compare the extent of the shell (outer, black circle), the ejecta `blob'
that contains the filaments (intermediate, white circle) and the condensation centered on the nucleus (smallest, 
white circle). Each panel is $2.1^\circ$ ($1.4\times 10^7$ km or 0.09 AU) on a side.
\label{holmesreg}}
\end{figure}

Figure~\ref{holmesreg} compares the 24 $\mu$m image to the optical image taken closest in time. 
The infrared image is significantly more sensitive, so the outermost extent of the shell is larger in that image;
however the brightness profiles of the shell are compatible. 
On the other hand, the brightness distribution closer to the nucleus is noticeably different. While the `blob' has
approximately the same location and shape, the central condensation near the nucleus is not present in the optical
image. Indeed only a faint nuclear source was evident at the location of the nucleus, with magnitude $15.5\pm 0.3$
measured in comparison to the 11.4 mag star HD 27590 and the 15.7 mag galaxy MCG+06-09-013. 
The total reflected light from the inner core is very small in comparison to the more extended emission.
At the
location of the nucleus, a compact enhancement of approximately 30\% in brightness is present, above the 
much more extended `blob' plus shell.
For comparison, in the infrared image, the peak brightness in the 
central core is 2 times {\it brighter} than the more-extended components.
This difference indicates different particle properties in the core, such that they are less effective at scattering
than the more extended emission. Indeed, this result obtained purely from the optical and infrared images could
be expected, to some extent, from the interpretation of the mid-infrared spectra. The core is due to much larger particles than
the extended shell; the shell particles radiate less efficiently at infrared wavelengths while they scatter
approximately as effectively as the larger particles at visible wavelengths.
The infrared surface brightness
scales as surface area times emission efficiency and blackbody function, while
the visible brightness scales as the surface area times scattering efficiency.
If the core particles are larger than the wavelength the 24 $\mu$m infrared wavelength, and the shell particles
are larger than the 0.6 $\mu$m visible wavelength, then the core/shell contrast in the infrared image
should be greater than that in the visible image by a factor
$Q_{IR}(a_s)^{-1} B_\nu(T_c)/B_\nu(T_s),$
where $Q_{IR}(a_s)$ is the absorption efficiency of shell grains at 24 $\mu$m wavelength, $B_\nu$ is
the Planck function, and $T_c$ and $T_s$ are the temperatures of core and shell grains, respectively.
The observed core/shell contrast in the infrared image is at least 6 times that in the visible image.
The relative lack of contrast of the core relative to the extended emission in the
optical image can be explained if the shell particle size is $\sim 1$ $\mu$m, similar to that
derived below (2 $\mu$m) based on the 2008 Mar image morphology.

\clearpage

\section{Model for the dynamical evolution of debris}

The images and spectra of the ejecta reveal details about the nature of the
explosion. We developed a numerical model for the debris evolution by adapting the model of \citet{vaubaillModel}.
Particles of a range of sizes, from 0.1 $\mu$m to 10 cm, with a power-law size distribution
with number density per unit radius $\propto a^{\alpha}$, are ejected 
over an interval $\tau_{exp}$. 
The velocity as a function of particle size
is taken from the hydrodynamic model of \citet{crifo95}. The direction and extent of the ejection was modeled as
a cone with opening half-angle $\Theta$; the angular distribution within the cone was modeled as
a power law $\cos^n \theta$ where $n=0$ means a uniform distribution across the cone.

\subsection{Hemispherical ejection models}
\begin{figure}
\epsscale{.5}
\plotone{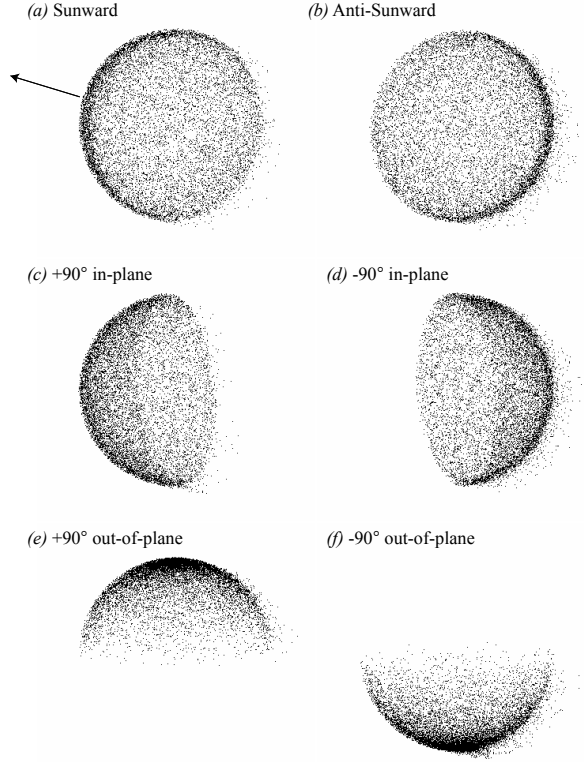}
\epsscale{1}
\figcaption{Hemispherical ejection models for comet Holmes on 2007 Nov 11. 
All particles were  ejected over a 1 hr period in these models, and 
all these models are identical, save
for the direction of the axis of the hemisphere into which the debris are strewn. For the Sunward and Anti-Sunward
models in panels {\it (a)} and {\it (b)}, the ejection is toward or away from the Sun. 
For the models in panels {\it (c)} and {\it (d)}, the
ejection is along an axis rotated $90^\circ$, in the orbit plane, from the sunward direction.
For the models in panels {\it (e)} and {\it (f)}, the
ejection is along an axis rotated $90^\circ$, out of the orbit plane, from the sunward direction.
Only the model in panel {\it (a)} produces the sunward limb-brightening of the shell as observed.
The images orientations are appropriate for observations from Earth, with north up and east to the left;
the projected vector to the Sun is $70^\circ$ E of N (as indicated by the
arrow in panel {\it (a)}.
\label{fig:direj}}
\end{figure}

The images, both in 1892 and in 2007, strongly suggest an explosion over a localized portion of the surface.
The observed morphology for the fastest-moving,
small particles is a shell, with significant limb-brightening in the sunward direction and a more open
shape in the antisunward direction.
A spherical ejection model yields a limb-brightened shell, but it cannot match the observed asymmetry.
Consider the models in Figure~\ref{fig:direj}. Each panel is a model with $\theta=90^\circ$ and $n=0$, i.e.
a uniform distribution of ejecta over a hemisphere. The edge-brightening is caused by a combination of
the shell being relatively empty plus projection effects on the line of sight.
The observations are only matched by a sunward ejection.

\subsection{Conical ejection models}

The hemispherical ejection models can match the morphology of the outer shell, but they do not
generate the pronounced peak in ejecta located just anti-sunward of the nucleus. This peak has
been referred to as the `false nucleus' or `pseudonucleus' or more prosaically, the `blob.' 
This feature was evident even to naked-eye
or binocular observers in 2007, and it is evident in the 1892 images and visual observation notes \citep{barnard13}.
We found that the anti-sunward peak is naturally produced by an explosive ejection of debris into a conical distribution
that is somewhat offset from the anti-solar direction. The parameter space for such models (longitude and latitude of
the cone, $\Theta$, $n$, in addition to the other parameters like size-distribution and ejection
history) is large and has not been exhaustively searched. But a plausible model was found by
experimenting with several possible values and comparing slices through the model image to
slices through the observed image. 

\begin{figure}
\epsscale{.5}
\plotone{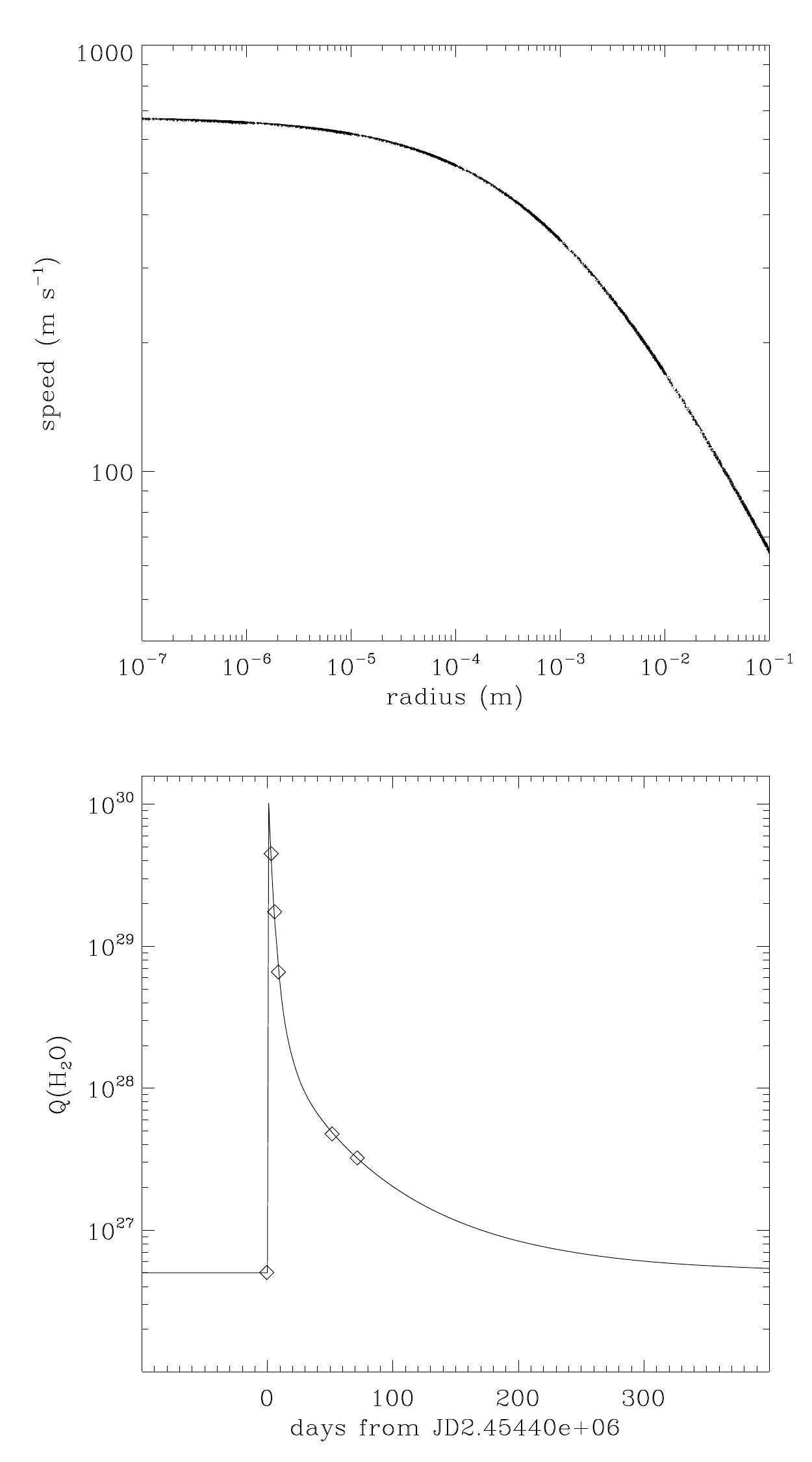}
\epsscale{1}
\figcaption{Parameters for the conical ejection model. {\it (top)} Ejection speed versus particle size.
{\it (bottom)} Molecule production rate versus time.
\label{fig:vr}}
\end{figure}

The ejection speeds for the particles are based on the \citet{crifo95} model; Figure~\ref{fig:vr} shows the speed
versus particle size. The dependence of velocity on sub-solar angle was not taken into account, so the ejection
is always radially outward from the nucleus. The velocity model was modified from the \citet{crifo95} model
so as to match the relative location of the `blob' with respect to the nucleus. Several models for the emission
history were considered, including a single impulse, natural production plus an impulse, and natural production
plus a decreasing excess. This latter model is able to reproduce the distribution of material plus the debris trail.
Figure~\ref{fig:vr} shows the production rate for the model together with the measured molecule production rate.

\begin{figure}
\epsscale{.5}
\plotone{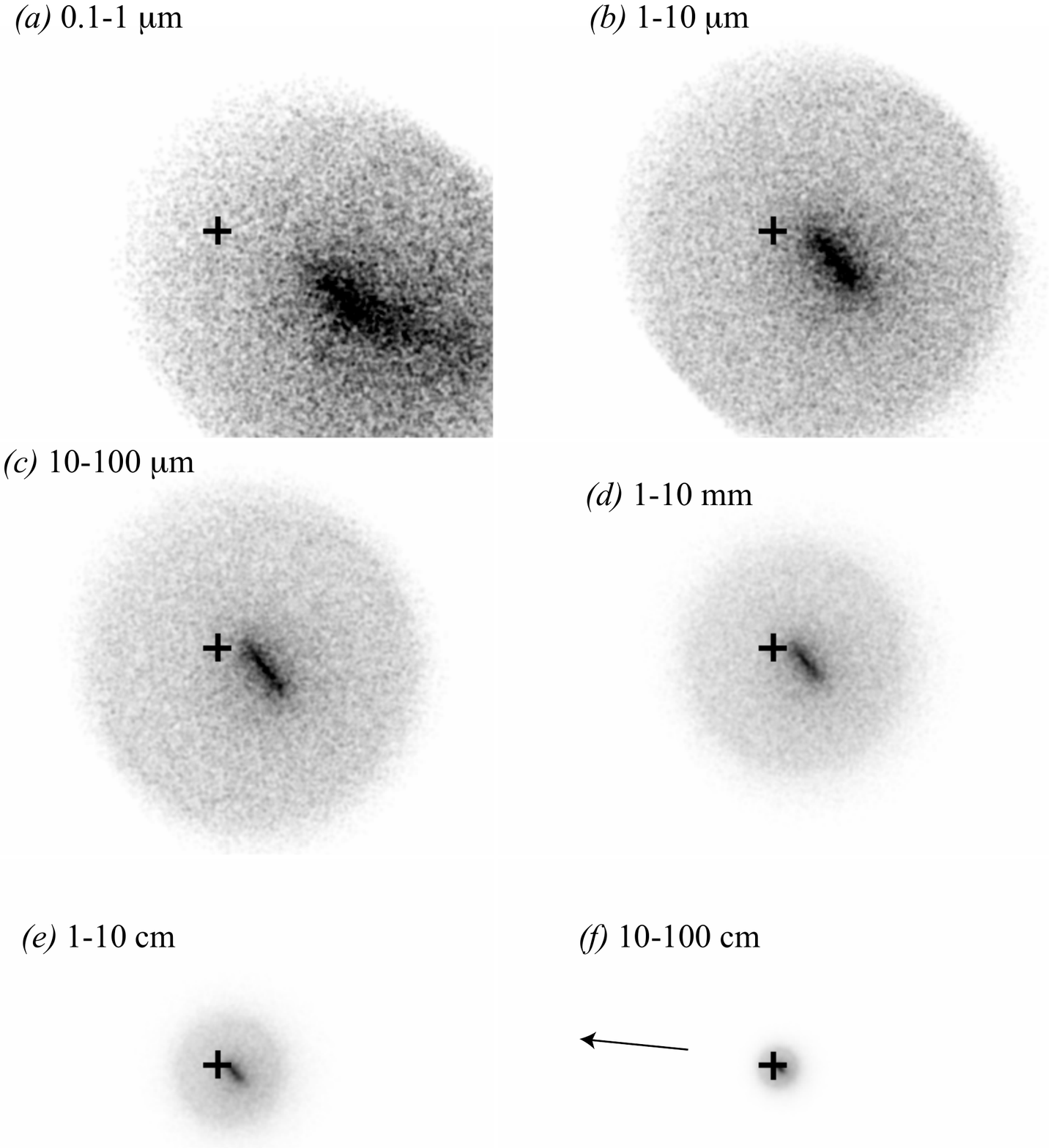}
\epsscale{1}
\figcaption{Conical explosion model for comet Holmes on 2008 Mar 13. Each panel shows the distribution of particles
in one logarithmic size bin, ranging from 0.1-1 $\mu$m {\it (a)} to 10--100 cm {\it (f)}. 
A black cross marks the location of the nucleus in each panel. 
In each panel, a similar morphology is evident, as it is in the observed image. There is a diffuse shell, with a 
linear ridge of enhanced brightness that begins at a location offset from the nucleus and extends generally in the
anti-sunward direction from the date of the explosion ({\it not} the anti-solar direction on the date of observation).
The ridge is a projection effect from the center of the ejection cone.
The smallest particles are fastest, so they are the only ones to reach the edge of the shell. 
The forward edge of the shell, and the location of the ridge of
enhanced brightness, becomes progressively closer to the nucleus for smaller as the particle size increases.
The offset between the peak column density for this model approximately matches the observed one for a
nominal size distribution. An arrow in panel {\it (f)} is $20^\prime$ long and points toward the Sun.
\label{fig:cone}}
\end{figure}

Figure~\ref{fig:cone} shows the model images for each of 6 logarithmic size bins. The offset of the peak of
emission with respect to the nucleus is produced by having a focused, conical ejection from a specific 
point on the nucleus.
To match the observed location of the `blob' of emission behind the nucleus, the origin of the explosion must
be offset from the subsolar point by $-20^\circ\pm 3^\circ$ in the orbit plane and $-3^\circ\pm 2^\circ$ perpendicular
to the orbit plane.
The opening angle that best matches the observations is $\Theta\sim 45^\circ$; however, note that this value is 
closely tied to $n$ as well as the ejection velocity.
The more sharply focused models, $n>0.75$, generate too much of a `blob,' relative to the shell. 
The uniform models produce a weak `blob.' The models that look most like the observed
images have $0.25<n<0.75$.

Our goal in generating the models was to qualitatively explain the image morphology, and to quantitatively
determine the range of particle sizes that dominates each portion of the image. The determination of the
particle size is critical for estimating the shell mass and kinetic energy, which place fundamental constraints on
the nature of the explosion and properties of the nucleus. We find that the outermost shell is dominated by
small particles, in the 0.1--10 $\mu$m range; we will adopt a nominal size of 2 $\mu$m for the shell. 
The `blob' is also due to small particles, in the 1--100 $\mu$m size range. The largest of the particles
produce a sharper filament at the center of the `blob.' The largest particles remain close to the nucleus and
could in principle explain the strong peak at the location of the cometary nucleus. 

The near-nucleus coma in 2008 Mar 13 is likely to have a contribution from both from large ejecta particles
dating back to the explosion as well as naturally-produced large particles. That the comet naturally
produces large particles is evident from the existence of its debris trail. This trail comprises mm-sized
particles from previous orbits of the comet, and it closely follows the comet's projected orbit. 
Smaller particles are blown away from the comet's orbit by radiation pressure.
Such debris trails are nearly always present around short-period comets \citep{reachtrail} and
are physically associated with meteoroid streams \citep{jenniskens}.
The influx of material from the 2007 perihelion passage is
expected to still be close to the nucleus, so that the portion of the trail that we observed in 2007 is
dominated by the previous perihelion passage. The large-particle production from the 2007 explosion is
thus difficult to assess. The image morphology, in particular the filaments seen in the {\it Spitzer}/MIPS images
 in 2008 Mar, seem to suggest an inner edge to the ejecta distribution $\sim 4'$ from the nucleus. Such an edge can only occur if there
 is an upper limit to the particle size distribution. Based on the location of the inner edge, the upper
 limit particle size is $\sim 2$ cm.

The model presented here only approximately matches the observed images; there is still significantly
more information in the observed images than the models can reproduce accurately. 
Some discrepancies between the cone model and the observed images include the lack of
a sharp forward edge to the shell, the size of the shell relative to the `blob,' the lack of a central coma near
the nucleus in 2008 Mar. Some of these discrepancies could be explained by combining a conical plus
hemispherical model and extending the time over which ejecta are launched.
None of these simple models produce structure in the `blob' that can be associated with the 
observed filamentary
features. 

\subsection{Multiple-explosion model}

The thin, filamentary features evident in spatially-filtered optical or infrared images may represent
multiple explosive events on the surface. The features persist over many months, with similar orientation,
and they grow approximately in proportion to the shape of the ejecta `blob.' 
While analogy to split comets initially led to suspicion that the filaments were large chunks of ejecta
that disintegrated after release, their distribution does not resemble that of split comets,
for which the fragments generally lie along the comet's orbit. Futhermore, the longevity and great size of the filaments
strongly suggests they are peaks in the debris column density rather than a small number of large fragments.
The filaments are more likely to represent either (1) separate holes in the nucleus, or
(2) multiple eruptions from the same hole. 
In both cases, the eruptions likely occurred on the same day, and all events were on the datyime side of the nucleus.
If the filaments are due to separate holes on the surface that erupted at the same time, then they subtended 
approximately $40^\circ$ across the nuclear surface. 
The rotation period of the comet before explosion was found
based on visible photometry to be in the range 7-12 hr, typical for a short-period comet \citep{snodgrass06}.
If the filaments are due to a single hole that erupted multiple times, then the interval between first and last
events was $\sim 3$ hr.

\section{Nature of the explosion}

\subsection{Determination of ejecta mass and energy\label{massenergy}}

Using the insights on particle size from the dynamical model and the spectra, 
we can now estimate the mass and kinetic energy of the ejecta.
We segment the 2008 Mar {\it Spitzer}/MIPS image into three regions: 
the {\it shell} which is the majority of the area subtended by the comet,
the {\it blob} which is the peak in brightness centered roughly anti-sunward of the nucleus, and
the {\it core} which is the peak in brightness centered on the nucleus.
Table~\ref{masstab} lists the observed and derived properties of these regions.
To calculate the mass, we assume each emitting region (indexed by $i$) 
is a uniform sphere with the specified average surface brightness and radius ($R_i$), 
and the grains in each region  are spherical with the specified size, $a_i$, temperature, $T_i$, and bulk 
density, $\rho=1$ g~cm$^{-3}$.
The grain size is smallest in the shell, and a nominal value of 2 $\mu$m was derived above from
the dynamics and the presence of a uniformly bright silicate feature in the 2007 Nov mid-infrared spectra.
We then compute the total flux density, $F_\nu^{(i)}$, of each component and solve for the mass
\begin{equation}
M_i = \frac{8}{9} \frac{ \rho a_i \Delta^2}{B_\nu(T_i) Q_i}F_\nu^{(i)},
\label{masseq}
\end{equation}
where $B_\nu$ is the Planck function,
 and $Q_i$ is the absorption efficiency of  grains of radius $a_i$ for photons of wavelength 24 $\mu$m. 
The Table was calculated for $Q_i=1$ for the core and blob but $Q_i=2\pi a/\lambda=0.52$ for the
shell where the particles are smaller than the wavelength.
To account for realistic mineralogy, we also used Mie theory to 
calculate $Q_i$ for grains composed of amorphous silicates, specifically
MgFeSiO$_4$ with optical constants from \citet{dorschner95}, and for amorphous C.
While we know the shell is composed of fine silicates (based on the 2007 Nov spectrum), the
composition of large particles in the core (or blob) of the 2008 Mar image is not known. 
Changing their composition from grey to amorphous silicates changes the
mass of the core, blob, and shell by +25\%, -32\%, and -40\%, respectively.
Changing their composition from grey to amorphous carbon changes the
mass of the core, blob, and shell by +59, -20\%, and -24\%, respectively.
The total mass, summing over components, is $9.9\times 10^{12}$ g, with comparable
contributions from all three regions.

To calculate the kinetic energy for each region, we use late-time (2008 Mar 13)
estimates (distance/age) for the separation of the blob from the nucleus,
the expansion velocity of the shell, and a rough estimate for the core.
The expansion speed of the coma was much higher at early times \citep[cf. 554 m~s$^{-1}$;][]{lin09},
but there was a significant contribution from spectral bands of  gas  which carry little mass.
The separation of the blob from the nucleus was steady (linear) over the entire period, with
measured speed $132\pm 4$ m~s$^{-1}$ on 2007 Nov 1 \citep{lin09} only slightly higher than our
estimate based on the 2008 Mar 13 infrared image.
The total kinetic energy is $1.2\times 10^{21}$ erg, being dominated by the shell, despite the small
size of the particles, due to the much higher expansion speed and $v^2$ dependence of the kinetic
energy. If the mass of the shell was conserved, and the shell decelerated from its early-time 
expansion speed, then the kinetic energy is 4.5 times larger than our conservative estimate.
For comparison to some other explosions, the kinetic energy of the ejecta from the 2007 Holmes event 
was at least 6300 times the kinetic energy of the impactor that the {\it Deep Impact} mission sent into comet Tempel 1,
and the Holmes ejecta kinetic energy is equivalent to 31 kTon of TNT.
The total energy of the event is certainly larger than the observed kinetic energy, but the non-kinetic energy remained
on the nucleus in the form of seismic waves and material deformation.

\begin{figure}
\epsscale{.8}
\plotone{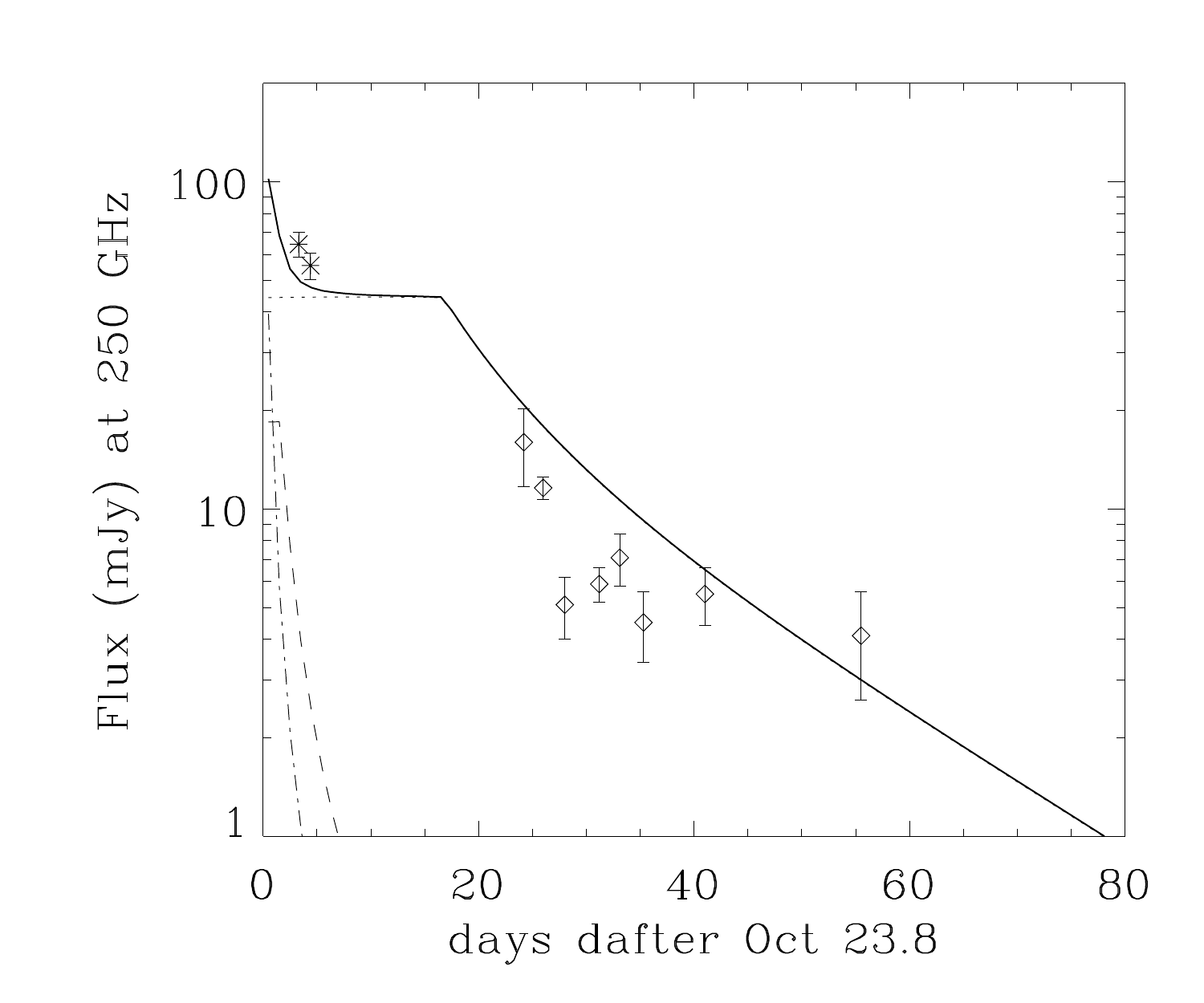}
\epsscale{1}
\figcaption{Evolution of the 250 GHz flux within an 11$''$ beam (diamonds) 
 and approximate equivalent flux determined from earlier
microwave observations (asterisks); from \citet{alten09}.
The solid curve shows the predicted flux based on the model developed from the MIPS 24 $\mu$m
image. The dotted, dashed, and dash-dotted lines are for the core, blob, and shell
components, respectively.
\label{alten}}
\end{figure}

Validation of this determination is possible, thanks to 250 GHz (1.2 mm wavelength) continuum
observations during the 2 months after the explosion \citep{alten09}. 
We used the core+blob+shell model that was derived from
the shape and brightness of the MIPS 24 $\mu$m in 2008 March image to predict the evolution of the 
250 GHz flux
within the fixed 11$''$ beam of the 30-m telescope, from 2007 Oct through 2008 Jan. 
Each of the three components (core, blob, shell) was
treated as an expanding spherical shell with mass, velocity, and 
particle size from Table~\ref{masstab}.
When the entire flux of a component of the comet is measured, then its flux density
is obtained by inverting equation~\ref{masseq} and using the absorption efficiency at
250 GHz for $Q_i$.
But at later times (once a component expands to a size larger than the beam of the 30-m telescope), the 250 GHz flux density of each component $i$ becomes
\begin{equation}
F_{250}^{(i)} = \frac{9}{8\pi} \frac{B_\nu(T_i)Q_{250,i} }{\rho a_i (v_i t)^2} M_i \Omega_{b},
\end{equation}
where $t$ is the time since the explosion on 2007 Oct 2, $\Omega_{b}$ is the
beam solid angle, and $v_i$ is the expansion velocity of component $i$.
The flux is initially a constant, then after the component is larger than the beam, its flux
decreases as the inverse square of the time since explosion.
Figure~\ref{alten} compares the model predictions to the observed data. 
After 2007 Nov 3, the 250 GHz flux is produced exclusively by the `core' component of the image.
This component was smaller than the beam of the 30-m telescope until 2007 Nov 10, after which
it became resolved and its flux decreased. The observations with the 30-m telescope began on 
2007 Nov 16, so they sample only the spatially-resolved part of the light curve. The agreement between
the model derived from the MIPS 24 $\mu$m image and the IRAM 250 GHz light curve is excellent:
it required no adjustment of parameters.
Two early microwave observations on 2007 Oct 27 at 88.6 GHz and 2007 Oct 28 at 90.6 GHz were
scaled to approximate the flux at 250 GHz \citep{alten09}. 
Our model does not match these two early observations. However, if we adjust the grain size in
the `blob' from 8 $\mu$m to 30 $\mu$m (and increase its mass proportionally, so that the MIPS flux
is preserved), then an excellent fit is obtained. Thus the early-time microwave observations probably
detect the largest particles in the blob, or the smallest particles in the core.
The mass derived from our model is 10 Tg (where Tg means $10^{12}$ g), 
which may increase to 15 Tg if
we included the modified model accounting for the early-time microwave observations. 

Our mass estimate is somewhat lower than some previous estimates.
Based on the 250 GHz data, \citet{alten09} derived a mass of 210 Tg.
Since our model matches the same 250 GHz data, the discrepancy may arise in a different assumed
grain size and production history; we assumed the particles were produced in an explosion rather
than continuously.
Based on the optical brightness shortly after the explosion, \citet{sekanina08} derived a mass of 100 Tg in 2 $\mu$m grains.
Based on a tentative detection of background starlight extinction, \citet{montalto08} derived a mass
of 1 to 100 Tg. Using our 3-component dust shell model, the predicted optical depth of the Holmes ejecta
at 4 days after the explosion is $8\times 10^{-4}$, which is too small to have been detectable via extinction
of starlight. If the extinction results are correct, they are difficult to reconcile with 
the infrared emission.

\clearpage

\subsection{Model for the explosion}

The massive, rapidly expanding clouds of debris found around comet Holmes in 1892 and 2007  are ejecta from an explosive event that occurred on the nucleus. The trigger for the explosion, and the source of the energy, are not known. A likely suspect for the energy source is the latent heat produced by crystallization of subsurface amorphous ice
\cite[][, and references therein]{prialnik02,prialnik04}.
Cryo-volcanism driven by amorphous ice has been applied to explain small outbursts of comet 9P/Tempel \citep{belton08}
as well as the smooth terrains on the surface of 9P/Tempel 1 \citep{belton09}.
We use the term `explosion' for the comet Holmes events in 1892 and 2007, to distinguish it from the much smaller events that
are commonly referred to as `outbursts' on other comets; the events may have a similar origin but are unlikely to
be identical in all regards due simply to the tremendous energy of the Holmes explosions.
Water condensed from vapor at low temperature takes on an amorphous form, and when heated above its glass transition temperature
$T_g\sim 140$ K, it undergoes an irreversible, exothermic phase transition \citep{mcmillan65}. The energy released by crystallization of amorphous ice, per unit mass, has been measured experimentally to be 
$\mathcal{L}\simeq$1200--1800 J/mole \citep{ghormley56,ghormley68}.
(If a different energy sources for the explosion is identified, it could be substituted in this model based on its
specific energy production rate, $\mathcal{L}$.)
The surfaces of short-period comets certainly reach temperatures above $T_g$ if their perihelia are smaller than 4 AU.
If a pocket of amorphous ice is warmed, by the gradual progression of heat into the nucleus after many perihelion passages, the latent heat will be released into the material surrounding the cavity. 

The cold, amorphous ice that was vapor-deposited from the early solar nebula 
(as opposed to the surface layers that have been exposed to the inner solar system)
contains trapped volatile gases
that are relics of the solar nebula composition at the time of the comet's formation. Upon heating above the glass transition temperature, the volatiles are released \citep{jenniskensblake96,barnun07}. Thus in addition to the latent heat, the vapors of the formerly trapped volatiles will be expanding from the former amorphous ice. 


\begin{figure}
\epsscale{1}
\plotone{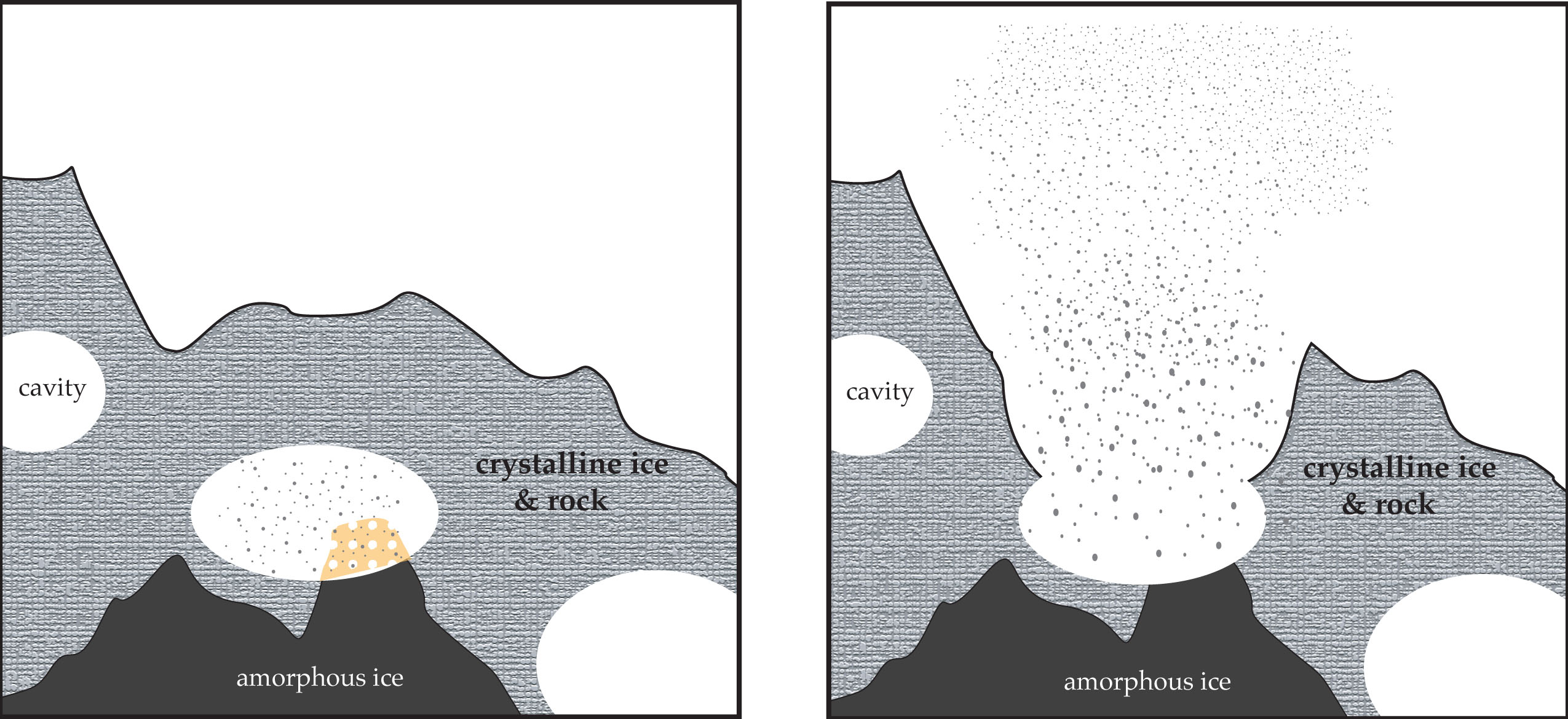}
\epsscale{1}
\figcaption{Internal structure of a comet that could lead to an explosion. The configuration before and during the explosion are shown in the
left and right panels, respectively. 
The pristine ice that formed from the solar nebula is gas-laden, amorphous ice. Cavities pervade the nucleus; they are realized here as elliptical holes but in reality they may be much more complicated topologies that formed from the accumulation of cometesimals while the comet was being assembled. If some of the amorphous ice 
is heated (by the gradually progressing heat from the Sun each perihelion passage) above its glass transition temperature, and if that
amorphous ice is in contact with a cavity, then its trapped volatiles will be released into, and will pressurize, the 
cavity. If the cavity is enclosed, then the combined outgassing, together with the release of latent heat from the amorphous-crystalline
phase transition, provide positive feedback as more and more amorphous ice is crystallized, until the reservoir of amorphous ice is depleted, 
the vapors can be vented to other cavities, or the pressure of the cavity exceeds the tensile strength of the 
crystalline ice and rock that form the surface material. For comet Holmes, the cavity
pressure exceeded the tensile strength, which was at least 2.7 kPa, and the vapors and surrounding material of the comet were accelerated to form the dramatic plume of ejecta with $10^{21}$ erg of kinetic energy. 
\label{holemod}}
\end{figure}

Let us consider the case where a pocket of amorphous ice crystallizes below the surface; gases trapped within the amorphous ice are released and 
other subsurface ices sublimate, leading to a buildup of gas pressure. 
Figure~\ref{holemod} illustrates a possible configuration for the various materials.
The pressure builds until it exceeds the
strength, $S$, of the nuclear material above the gases, which then erupt through the surface hole together with entrained solid
debris. Let us approximate the hole and subsurface region as a cylinder with radius $r_h$ and depth $d_h$.
The debris move together with the expanding gases for a distance $R_{dec}$ from
the nucleus, where the collision rate between debris and gas become too low and they decouple.
We equate the kinetic energy of the debris, $E_k$, with the work done on the debris by the expanding gases,
\begin{equation}
E_k = S \pi r_h^2 R_{dec},
\end{equation}
where we set the pressure at the surface of the hole equal to the strength of the material and we approximate the
force on the debris being constant from the time of release from the surface until decoupling.
We equate the observed mass of debris, $M_d$, with the mass of material in the hole,
\begin{equation}
M_d = \pi r_h^2 d_h \rho_h,
\end{equation}
where $\rho_h$ is the mass density of the nucleus.

This model allows us to take the observed properties of the ejecta, specifically its mass and kinetic energy, and 
constrain the surface properties of the comet and the energy source of the explosion.
Using the observed kinetic energy,  $E_k=1.2\times 10^{21}$ erg, and $\mathcal{L}=1200$ J/mole,
we infer that $2\times 10^{12}$ g of amorphous ice crystallized. 
Taking the decoupling radius to be 10 times the nuclear radius, $R_{dec}=10 R_N$, with
$R_N=1.7$ km \citep{LamyNuc}, we can constrain the product of hole area and surface strength.
First, if the material strength of the nucleus is $S=200$ kPa, 
which is less than that of pure water ice (1000 kPa) but
more than lunar regolith \citep[1 kPa; see][]{holsapple07}, then the hole has radius 115 m.
Second, if we assume a hole radius 1000 m, similar to the largest topographic features on the nucleus
of comet Wild 2 \citep{brownlee06}, then the surface strength is 2.7 kPa; this strength estimate
is probably a lower limit, since a much larger `hole' would occupy much of the comet's surface and would have
thrown ejecta into a wider solid angle than the images indicate.
Thus for the plausible range of surface strengths 2.7--200 kPa, the required hole radius is 1000--115 m, respectively;
to match the observed mass, the depth of the hole for these models is 4.4--330 m, respectively (assuming the mass density
in the nucleus is $\rho_h=0.6$ g~cm$^{-3}$).
For all models in this range, the volume of the nucleus that was expelled from the comet is about twice the volume of the amorphous ice that crystallized to power the explosion.

The effect of the explosion on the nucleus depends on the internal structure of the nucleus.
Our understanding of the surface of cometary nuclei is very limited;
its outer layers are considered to comprise a rubble mantle of devolatilized rocks too large ($\gg$ cm size)  to have
been levitated from the surface by drag from outgassing \citep{jewitt05}, 
though the results of the {\it Deep Impact} as requiring fine dust ($\sim 2$ $\mu$m size) to a depth of tens of meters
\citep{ahearnT1}.
The range of possibilities based on observation and theory could reflect comet-to-comet variations due to the
range of physical conditions and compositions in the outer solar nebula, or they could be based on surface variations
with icy surface or just-subsurface regions occupying a small or temporally varying portion of the nucleus.
In general terms, stress from the explosion would damage the nucleus by driving cracks through it, and
the energy of the explosion is large enough that the body should be significantly damaged unless its tensile strength over 
scales $>100$ m is large.
Some guidance as to the implied strengths and the capability of the nucleus to survive disruption can be obtained from studies of the effects of collisions.
For monolithic bodies, the catastrophic disruption threshold, determined by N-body simulation, ranges from 200--$10^4$ J~kg$^{-1}$ depending
on the strength model, with the lower threshold corresponding to a tensile strength of 100 kPa  \citep{stewart09}. 
Using the observed kinetic energy of the Holmes ejecta, and the mass of its nucleus (for 0.4 g~cm$^{-3}$ bulk density),
the specific energy for the 2007 explosion was 6 J~kg$^{-1}$. 
The specific energy for the 2007 Holmes explosion was thus close to that required to completely disrupt the nucleus,
but its survival is consistent with a tensile strength of 100 kPa or perhaps as low as 10 kPa though
that is outside the range of models.

The effect of the explosion is somewhat different in the case of a `rubble pile' construction, in which the nucleus is comprised of blocks of material 
that are primarily bound by gravity \citep{weissman04}. 
Numerical results indicate rubble piles survive disruption from collisions with specific energies smaller than $\sim 2$--100 J~kg$^{-1}$ 
\citep{asphaug09}, i.e. a smaller energy per unit mass is required to disrupt a rubble pile.
If we treat the explosion as occurring in one block of the rubble pile, and that block has momentum equal and opposite to that
of the ejecta, then we find the closest analogous impactor properties; for blocks of radius 100, 200, 500, 1000 m, the
speeds are 170, 22, 1.4, and 17 m~s$^{-1}$. The specific energy of the `impact' from such blocks is 27, 3.4,
0.2, and 0.02 J~kg$^{-1}$, and their mass ratio to the nucleus is 0.00018, 0.00114, 0.023, and 0.18.
Comparing to the rubble pile impact models of \citet{asphaug09}, explosions into blocks with size greater than 200 m radius
have low enough specific energy that they would not disrupt the nucleus, while an explosion in 100 m block would drive that block
into the nucleus with enough energy to disrupt it.
The effects of an explosion on a  comet nucleus may be rather different from those of a collision, especially in the case of a rubble
pile construction and heterogeneous composition.  
Taking into account just the gravitational potential energy of subsurface material displaced by the sinking block,
a block of 500 m radius or larger would be driven inward but would not reach the center, while smaller blocks would penetrate the nucleus
and be ejected from the other side.

\section{Comparison of 1892 and 2007 explosions}

\begin{figure}
\epsscale{.5}
\plotone{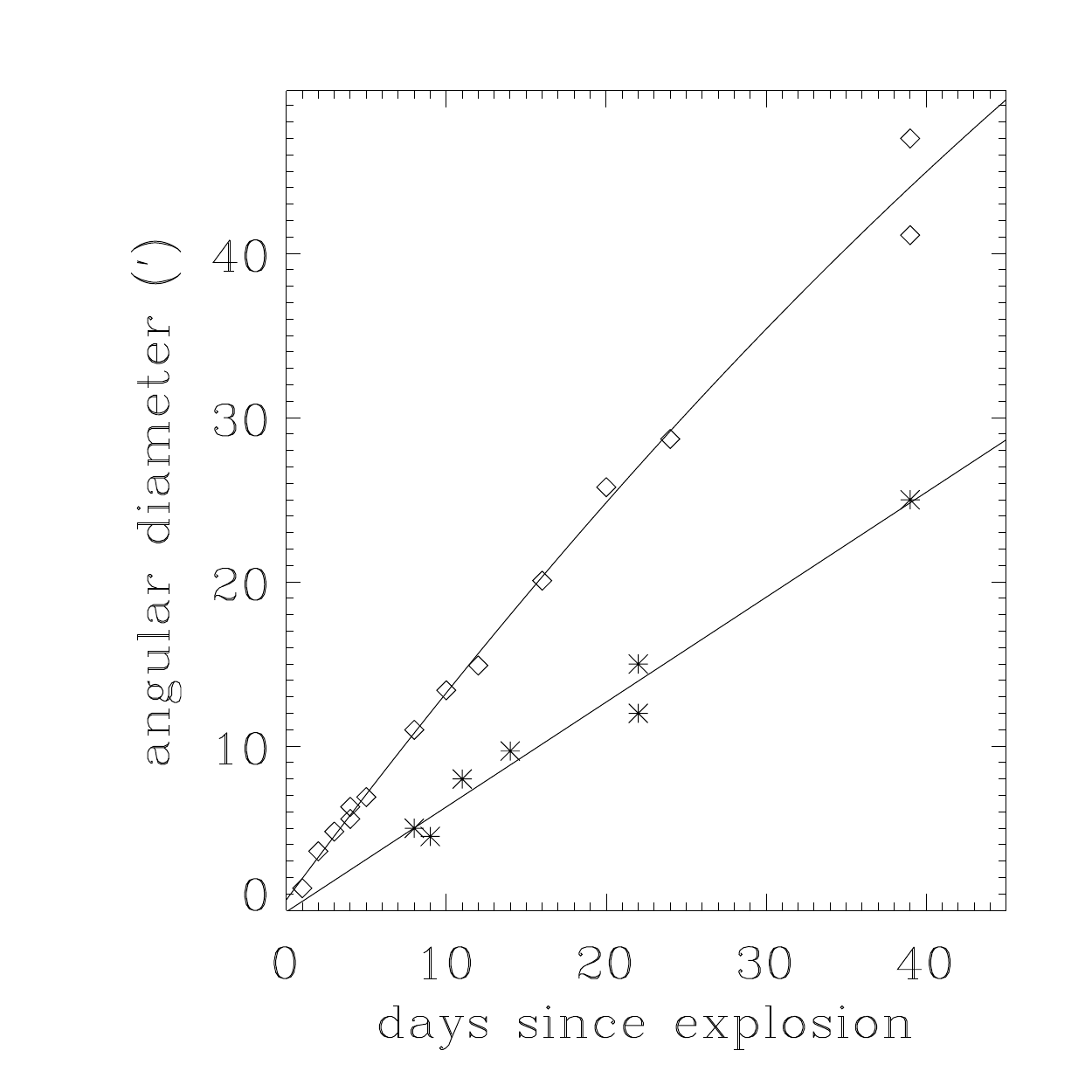}
\epsscale{1}
\figcaption{The expansion of the debris clouds from the 1892 ($*$) and 2007 ($\diamond$) explosions. The 2007 expansion
was somewhat nonlinear; the decelerating expansion was fit with a quadratic. 
No deceleration was evident in 1892, though the data are probably too sparse too show it.
\label{diamplot}}
\end{figure}

The appearance of the comet after its 1892 explosion is remarkably similar to its appearance after its 2007 explosion.
We performed a careful comparison using photographs obtained by \citet{barnard13}, as scanned by
J. McGaha and made available on the {\it International Comet Quarterly}\footnote{\tt http://www.cfa.harvard.edu/icq/17P\_outburst.html}. 
The scanned images were processed using {\tt astrometry.net} tools \citep{hoggastrom,langthesis}
in order to determine the location and orientation.
Table~\ref{obslogopt} lists the observing circumstances for the 1892 and 2007 optical observations, obtained using
the appropriate orbit of the comet from Horizons\footnote{\tt http://ssd.jpl.nasa.gov/?horizons}. The expansion of the debris cloud can be
approximately tracked by the increase in angular diameter with time; extrapolating back to zero angular diameter
suggests the 1892 explosion occured on 1892 Oct 31 ($\pm 1$ day). Figure~\ref{diamplot} shows the diameter
versus date since explosion for the 1892 and 2007 events. It is evident that
the 2007 explosion was more powerful than the 1892 explosion.
The debris expanded more rapidly in 2007, by a factor of 2.
The mass of debris cannot be accurately compared, for lack of infrared observations in 1892.
Comparing the optical images, based on the brightness of the debris compared to nearby stars, it appears
that the light reflected from the debris was somewhat fainter in 1892 than in 2007, by approximately a factor of 5 ($\sim 2$ magnitudes).
The energy of the explosion, as measured by the kinetic energy of the debris, was thus smaller in 1892 than in 2007 by a factor of $\sim 20$.

\begin{figure}
\epsscale{1}
\plotone{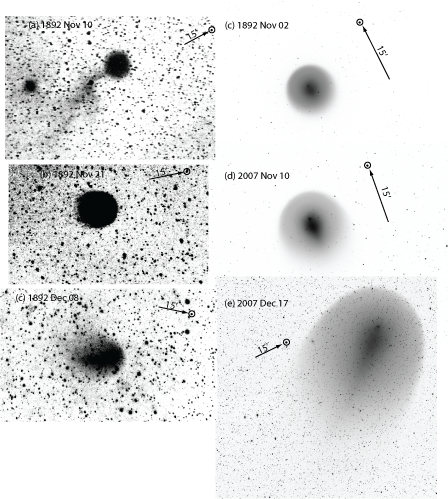}
\epsscale{1}
\figcaption{Comarison of optical images of comet Holmes in 1892 and 2007.
The 1892 images are from \citet{barnard13}; panels (a), (b), and (c) are scanned from photographs 
taken 11, 22, and 39 days after the 1892 explosion.
The 2007 images are from Holloway Comet Observatory; panels (d), (e), and (f) were taken
10, 18, and 55 days after the 2007 explosion.
All images have N up and E to the left. An arrow on each panel is 15$'$ long and points away toward the sun.
The 1892 Nov 21 photograph is heavily saturated, so no image detail except the outline can be discerned. 
The 1892 Nov 10 photograph is overexposed in the core but reveals faint widespread emission.
The faint emission to the SE, which \citet{barnard13} refers to as the ``attendant mass'', is almost certainly a cloud
of gas from the initial explosion. Remarkably, such a feature was also seen in the 2007 explosion in the
SuperWASP images on 2007 Nov 14.
\label{comp1892}}
\end{figure}

Figure~\ref{comp1892} compares images from the 1892 and 2007 events. 
The similarity between the images is clear, even though the 1892 observations are scanned pages from an
aged journal. A notable aspect in both cases is one relatively sharp rim of the expanding shell, with relatively
diffuse emission on the opposite side of the sharp rim. 
The description by \citet{barnard13} of the comet being `very diffused on the following side, but fairly sharply
defined preceding' agrees qualitatively with the appearance of the comet in 2007.
The sharp rim was directed toward the Sun in {\it both} the 1892 and 2007 events. 
In his descriptions of his photographs and visual observations, \citet{barnard13} refers to
a `preceding' and `following' side of the debris cloud, with the `preceding' side being sharper. From inspecting his
photographs to CCD images taken in 2007 and our own visual observations, it seems clear that he was calling
the sharper side the `preceding' one as a matter of convention (because the sharper side does not correspond to the
on-sky motion of the comet).

In addition to the expanding shell and rim, a relatively bright condensation of debris located just anti-sunward of the nucleus is present in both the 1892 and 2007 events. The ejecta condensation is visible in the 1892 Dec 8 image from \citet{barnard13} and in most optical images taken of the 2007 event. The brightness of the ejecta condensation may have been somewhat greater, relative to that of the rim and shell, in the 1892 event. The nonlinearity of the photograph makes it difficult to quantify this assessment, though inspection of Figure~\ref{comp1892} shows the comparison as clearly as possible.

\clearpage

\section{Conclusions}

Images of the ejecta from the 2007 Oct explosion of comet Holmes 
reveal a concentration that gradually moves away from the nucleus as well as a faster-moving 
hemispherical shell. Most of the kinetic energy is in the faster-moving dust in the shell, but most of the mass is in
the slower-moving particles closer to the nucleus. The mineralogy of the small particles seen in 2007 Nov indicates they are primarily
olivine and amorphous silicates; the mineralogy is similar to the ejecta from the {\it Deep Impact} experiment on comet Tempel 1.
Spectroscopy of the ejecta in 2008 Feb reveals only large particles, since the smaller particles have expanded far from the nucleus.
We derived properties of
 the ejecta including particle size and speed in order to match constraints from the infrared spectra in 2007 Nov and 2008 Feb,
and the brightness and morphology of the infrared image in 2008 Mar; the same model matches the temporal evolution of the
mm-wave flux of the comet observed from the ground.

We modeled the explosion as being due to a patch of volatile-laden amorphous ice being heated within an enclosure,
until the pressure within the enclosure exceeded the strength of the surrounding material. 
The inferred strength of the nuclear material $> 3$ kPa, with
a plausible value of 200 kPa. The relatively great tensile strength of the nucleus of comet 17P/Holmes, compared to
that estimated for other comets, for comparison 0.01-1 kPa inferred for the nucleus of 9P/Tempel 1 based on its mini-outbursts \citep{belton09},
may explain why comet Holmes has such exceptional outbursts. 

The 2007 explosion of comet Holmes was one of the most dramatic natural events produced by a comet
in recorded history, only clearly exceeded by
the impact of comet Shoemaker-Levy 9 on Jupiter's atmosphere in 1994 and the impact of a comet or asteroid 
on the Earth's atmosphere in 1908 near Tunguska. The kinetic energy of the debris from comet Holmes was $1.2\times 10^{21}$ erg,
about one order of magnitude less than the total energy the Tunguska impactor brought to Earth. For comet Holmes, the explosion
occurred with no outside perturbation: there is no evidence for an impact, and in any event it would be exceptionally difficult to
explain both the 2007 and 1892 outbursts using impacts, without seeing many comparable
events from other comets that traverse the same region of the solar system. 
That comet Holmes survived the two largest outbursts (in 2007 and 1892)
ever witnessed on a comet outside a planetary atmosphere indicates that this comet has significant internal strength
as well as internal structure.

The existence of an internal explosive energy source in comet Holmes, possibly driven by heating of a subsurface
pocket of amorphous ice, is relevant to other cometary and possibly asteroidal phenomena. Activity of comets, at
distances so far from the Sun that sublimation of common volatiles
cannot occur, is evidence for the amorphous-crystalline transition
driving activity \citep{meech09}. 
Repetitive outbursts discovered in the detailed observing campaign for comet 9P/Tempel 1 to support
the Deep Impact mission have been explained with fluidized flows of material driven by
subsurface amorphous ice \citep{belton09}, which is closely related to though much more gradual
than the mechanism we invoke to explain the comet Holmes explosion.
The frequent splitting of cometary nuclei \citep{jewittsplitting} remains unexplained.
If pockets of amorphous ice are heated and liberate $10^{21}$ erg of energy, such as we observed from the Holmes
ejecta, there is more than enough energy to drive loosely bound subunits of cometary nuclei apart at their observed
separation speeds, with total fragment kinetic energies of order or less than $10^{19}$ erg 
\citep[cf.][]{reach09sw3}. The freshly 
exposed surfaces of split comets, may be more likely to have explosive outbursts of the type exhibited
by comet Holmes in 1892 and 2007. Indeed, fragments from the interiors of split comets may have very short lifetimes
for disruption due to explosive phase transitions of exposed or near-surface amorphous ice. 
Comets making their first passages into the inner Solar System may have relatively more pristine material on their
surface, driving substantially higher activity rates.
Many of the brightest comets may have been outbursts, such as C/Hale-Bopp \citep[cf.][]{mccarthy07}. 
But older comets, on orbits that have kept them in the inner Solar System for $>10$ revolutions, may lose
a significant amount of their surface due to sublimation on multiple perihelion passages, insolation may eventually
reach their amorphous ice layer resulting in erratic and possibly eruptive activity similar to that of comet Holmes.
Activation of pockets of subsurface ice could possibly explain impulsive mass loss from the comets in 
the asteroid belt \citep{hsiehjewitt06}, though this application is less likely to apply in the same manner
as for cometary outbursts because asteroidal ice was not formed under outer solar system conditions that
led to the unstable, exothermic phase transition of amorphous ice that led to the explosion of comet Holmes.

\acknowledgements 

This work is based on observations made with the Spitzer Space
Telescope, which is operated by the Jet Propulsion Laboratory, California
Institute of Technology under a contract with NASA. Support for this work was
provided by NASA through an award issued by JPL/Caltech.

\clearpage

\bibliography{wtrbib}

\begin{deluxetable}{lrrccccccl}
\tablecaption{Log of {\it Spitzer} observations\label{obslog}}
\tablewidth{7truein}
\tabletypesize{\footnotesize}
\tablehead{
\colhead{Date} & \colhead{$R$} & \colhead{$\Delta_{S}$}
&  \colhead{Instrument} & \colhead{Wavelengths} & \colhead{Image Size} & \colhead{Comment}\\
& (AU) & (AU)\tablenotemark{a} && ($\mu$m) & ($\arcmin\times \arcmin$) &
}
\startdata
2007 Nov 10.82  & 2.505 & 1.866 & IRS  & 9.9--19.6 $\mu$m   & $0.49\times 0.20$ & Spectral map (short-high)\\
2007 Nov 10.82  & 2.505 & 1.866 & IRS  & 18.7--37.2 $\mu$m  & $0.59\times 0.37$ & Spectral map (long-high)\\
2007 Nov 10.82  & 2.505 & 1.866 & IRS  & 5.2--14.5 $\mu$m   & $0.91\times 0.63$ & Spectral map (short-low)\\
2007 Nov 10.82  & 2.505 & 1.866 & IRS  & 14.0--38 $\mu$m    & $2.52\times 0.74$ & Spectral map (long-low)\\
2007 Nov 10.85  & 2.505 & 1.866 & IRS  & 16 \& 22 $\mu$m  &             & Peakup image (saturated)\\
2008 Feb 27.16  & 2.964 & 2.342 & IRS  & 9.9--19.6 $\mu$m   & $0.49\times 0.20$ & Spectral map (short-high)\\
2008 Feb 27.16  & 2.964 & 2.342 & IRS  &  18.7--37.2 $\mu$m  & $0.59\times 0.37$ & Spectral map (long-high)\\
2008 Feb 27.16  & 2.964 & 2.342 & IRS  & 5.2--14.5 $\mu$m   & $0.91\times 0.63$ & Spectral map (short-low)\\
2008 Feb 27.16  & 2.964 & 2.342 & IRS  & 14.0--38 $\mu$m    & $2.52\times 0.74$ & Spectral map (long-low)\\
2008 Feb 27.19  & 2.964 & 2.342 & IRS  & 16 \& 22 $\mu$m  &   $1.41\times 1.03$    & Peakup image\\
2008 Mar 13.24  & 3.029 & 2.581 & MIPS & 24 \& 70 $\mu$m  & $120\times 120$ & Scan map\\
2008 Oct 25.23  & ...\tablenotemark{b} & ...\tablenotemark{b} & MIPS & 24 \& 70 $\mu$m  & $120\times 120$ & Scan map (shadow)\\
\enddata
\tablenotetext{a}{Distance from the observatory ({\it Spitzer}) to the comet}
\tablenotetext{b}{Shadow observation of the field where the comet had been, on 2008 Mar 13.24}
\end{deluxetable}

\begin{deluxetable}{lrrlcccrl}
\tablecaption{Log of optical observations used in this paper\label{obslogopt}}
\tablewidth{7truein}
\tabletypesize{\small}
\tablehead{
\colhead{Date} & \colhead{$R$} & \colhead{$\Delta$}
&  \colhead{Filter} & Pixel & \colhead{Diam.} & \colhead{$T-T_0$\tablenotemark{a}} & \colhead{Sun\tablenotemark{b}}& \colhead{Observer} \\
& (AU) & (AU) &  & ($\arcsec$) & ($\arcmin$) & (days) & ($^\circ$)\\
}
\startdata
1892 Nov 07	& 2.39 & 1.51 & Eye  & & 5    & 8 &  121 &Holmes\tablenotemark{c}\\
1892 Nov 08 & 2.40 & 1.52 & Eye      && 4.5 & 9 &  119 &Barnard\tablenotemark{d}\\
1892 Nov 10 & 2.40 & 1.53 & Phot.    &7.0& 8    & 11 & 116 &Barnard\tablenotemark{d}\\
1892 Nov 13 & 2.41 & 1.56 & Eye       && 9.7 & 14 & 111&Barnard\tablenotemark{d}\\
1892 Nov 21	& 2.44 & 1.63 & Phot.  &7.4& 12 & 22& 100 & Barnard\tablenotemark{d}\\
1892 Nov 21	& 2.44 & 1.63 & Eye    && 15 & 22& 100 & Barnard\tablenotemark{d}\\
1892 Dec 08	& 2.50 & 1.81 & Phot.  &10.7&  25 & 39& 84 & Barnard\tablenotemark{d}\\
\\
2007 Oct 27      & 2.45 &  1.63 & RGB & 2.13 &    &         4& 219 & Holloway\\
2007 Oct 28      & 2.45 &  1.63 & RGB & 2.13 &   &         5& 218 & Holloway\\
2007 Oct 29      & 2.46 &  1.63 & RGB & 2.13 &   &         6& 216 & Holloway\\
2007 Oct 30      & 2.46 &  1.63 & RGB & 2.13 &   &         7& 215 & Holloway\\
2007 Nov 02      & 2.47 &  1.62 & RGB &2.13 &    &      10& 210 & Holloway\\
2007 Nov 04      & 2.48 &  1.62 & RGB &2.65 &    &       12& 207 & Holloway\\
2007 Nov 05     & 2.48 &  1.62 & RGB & 2.65 &   &        13& 205 & Holloway\\
2007 Nov 06	& 2.49 & 1.62 & RGB & 2.65 &17 & 	14 & 204 & Holloway\\
2007 Nov 06     & 2.49 & 1.62 & r        & 2.65 &15 & 	14 & 204 & Reach\\
2007 Nov 08      & 2.50 &  1.62 & RGB &2.65 &    &       16& 199 & Holloway\\
2007 Nov 09     & 2.50 &  1.62 & RGB & 2.65 &   &        17& 197 & Holloway\\
2007 Nov 10     & 2.50 &  1.62 & RGB & 2.65 &   &         18& 195 & Holloway\\
2007 Nov 16     & 2.53 &  1.64 & RGB & 2.13 &   &        24& 176 & Holloway\\
2007 Nov 18     & 2.54 &  1.64 & RGB & 2.13 &   &        26& 176 & Holloway\\
2007 Nov 20	& 2.54 & 1.64 & RGB & 2.13 &25 &  28 & 173 & Holloway\\
2007 Dec 17     & 2.66 &  1.82 & RGB &4.26 &    &        55& 113 & Holloway\\
2007 Dec 26     & 2.69 &  1.91 & RGB &4.26 &    &        64& 101 & Holloway\\
2008 Jan 03	& 2.73 & 2.00 & RGB & 2.13 &52    & 72 & 94 & Holloway\\
2008 Mar 05     & 2.99&  2.99 & V      &  11.30 &  &        133& 75 & Holloway\\
2008 Mar 10     & 3.01 &  3.07 & RGB &10.33 &    &       138& 74 & Holloway\\
\enddata
\tablenotetext{a}{Time since explosion}
\tablenotetext{b}{Position angle (E of N) of the projected Comet-Sun vector at the time of observation}
\tablenotetext{c}{Telegram described by \citet{barnard13}}
\tablenotetext{d}{Visual observations and photographs from \citet{barnard13}}

\end{deluxetable}

\begin{deluxetable}{lllll}
\tablecaption{Mineralogical model for Holmes dust \label{comptab}}
\tablewidth{7truein}
\tablehead{
\colhead{Mineral} & \colhead{Abundance} & \colhead{formula} 
}
\startdata
amorphous carbon  & 5.63 & C\\
ferromagnesian sulfide  & 1.02 & MgFeS\\
amorphous pyroxene & 0.50 & (Fe,Mg)SiO$_3$\\
forsterite  & 0.46  &                Mg$_2$SiO$_4$\\
water ice   & 0.44 & H$_2$O\\
diopside  & 0.11 &  MgCaSi$_2$O$_6$\\
enstatite & 0.14 &  MgSiO$_3$ \\
water gas & 0.09 & H$_2$O \\
smectite & 0.06 &($\frac{1}{2}$Ca,Na)$_{0.33}$(Mg,Fe)$_3$(Si,Al)$_4$O$_{10}$(OH)$_2$ $\cdot n$H$_2$O \\
PAH  & 0.006 & C$_n$H$_m$\\
\enddata
\end{deluxetable}

\def\extra{
first set of numbers on word file
Composition & Abundance \\
amorphous pyroxene      &   0.33\\
smectite          & 0.12\\
forsterite        & 0.20\\
diopside      	 &  0.07\\
orthoenstatite  &    0.09\\
MgFeS       &   0.19\\
PAH      	  & 0.01\\
Water Gas   &   0.09\\
AmCarb       &  0.27\\
Water Ice     & 0.08\\}

\def\extra{looks lik eht eright set of numbers from latest word file in July 09
\begin{deluxetable}{ll}
\tablecaption{Mineral abudnaces in the Holmes ejecta\label{comptab}}
\tablehead{\colhead{Composition} & \colhead{Abundance} }
\startdata
amorphous pyroxene      &   0.50\\
smectite          & 0.06\\
forsterite        & 0.46\\
diopside      	 &  0.11\\
orthoenstatite  &    0.14\\
MgFeS       &   1.02\\
PAH      	  & 0.006\\
AmCarb       &  5.63\\
Water Ice     & 0.44\\
\enddata
\end{deluxetable}
}

\begin{deluxetable}{lllll}
\tablecaption{Derivation of Mass and Energy on 2008 Mar 13\label{masstab}}
\tablewidth{7truein}
\tablehead{
\colhead{Quantity} & \colhead{Core} & \colhead{Blob} & \colhead{Shell} & \colhead{unit} 
}
\startdata
Radius	        & 1.0 & 10 & 28 & arcmin\\
Radius         & 1.1 & 11.2 & 31.4 & $10^5$ km\\
Brightness	  & 45  & 10  &18 & MJy~sr$^{-1}$\\
Flux             & 17.2 & 383 & 5403 & Jy\\
Expansion speed	& 9 & 93 & 262 & m~s$^{-1}$\\
\hline
Grain size    & 200 & 8  & 2 & $\mu$m\\
Grain temperature & 185 & 200 & 400 & K\\
Mass           & 3.9 & 2.7 & 3.3 & $10^{12}$ g\\
Kinetic energy & 0.002 & 0.12 & 1.1 & $10^{21}$ erg\\
\enddata
\end{deluxetable}

\end{document}